\documentclass[groupedaddress,superscriptaddress,aps,prd,preprintnumbers,onecolumn,11pt,fleqn,nofootinbib,nobibnote,eqsecnum,floatfix,showpacs]{revtex4-1}

\usepackage{graphicx}
\usepackage{bm,amsmath,amssymb}
\usepackage[mathscr]{eucal}

\long\def\comment#1{ }
\newcommand{\eqn}[1]{Eq.~\eqref{#1}}
\newcommand{\beq}{\begin{equation}}
\newcommand{\eeq}{\end{equation}}
\newcommand{\nn}{\nonumber\\}
\newcommand{\dif}{{\rm d}}
\newcommand{\rmd}{{\rm d}}
\newcommand{\rme}{{\rm e}}
\newcommand{\rmi}{{\rm i}}
\newcommand{\rmP}{{\rm P}}

\newcommand{\rmtr}{{\rm tr}}

\newcommand{\rmI}{{\rm I}}
\newcommand{\del}{\partial}

\newcommand{\order}[1]{\mcal{O}{(#1)}}
\newcommand{\mcal}{\mathcal}

\newcommand{\bk}{\bm{k}}

\newcommand{\bp}{\bm{p}}
\newcommand{\bx}{\bm{x}}
\newcommand{\by}{\bm{y}}

\newcommand{\bz}{\bm{z}}
\newcommand{\bw}{\bm{w}}
\newcommand{\br}{\bm{r}}

\newcommand{\abar}{\bar{\alpha}_s}
\newcommand{\bbar}{b }

\begin{document}

%\preprint{arXiv:14mm.nnnn}

\title{Running coupling effects in the evolution of jet quenching}

\author{E.~Iancu}
\email{edmond.iancu@cea.fr}
\affiliation{Institut de Physique Th\'{e}orique de Saclay,
F-91191 Gif-sur-Yvette, France}

\author{D.N.~Triantafyllopoulos}
\email{trianta@ectstar.eu}
\affiliation{European Centre for Theoretical Studies in Nuclear Physics and Related Areas (ECT*)\\
and Fondazione Bruno Kessler, Strada delle Tabarelle 286, I-38123 Villazzano (TN), Italy}

\date{\today}

\begin{abstract}
We study the consequences of including the running of the QCD coupling in the 
equation describing the evolution of the jet quenching parameter $\hat q$ in the double logarithmic
approximation. To start with, we revisit the case of a fixed coupling, for which we obtain exact
solutions valid for generic values of the transverse momentum (above the medium saturation
scale) and corresponding to various initial conditions. In the case of a running coupling,
we construct approximate solutions in the form of truncated series obtained via successive
iterations, whose convergence is well under control. We
thus deduce the dominant asymptotic behavior of the renormalized $\hat q$ in the limit
of a large  `evolution time' $Y\equiv\ln(L/\lambda)$, with $L$ the size of the medium and
$\lambda$ the typical wavelength of a medium constituent. We show that the asymptotic
expansion is universal with respect to the choice of the initial condition at $Y=0$ and, moreover, 
it is remarkably similar
to the corresponding expansion for the saturation momentum of a `shockwave' (a large nucleus).
As expected,
the running of the coupling significantly slows down the increase of $\hat q$ with $Y$ in the
asymptotic regime at $Y\gg 1$. For the phenomenologically interesting value $Y\simeq 3$, 
we find an enhancement factor close to 3, independently of the initial condition and for
both fixed and running coupling. \end{abstract}

%\keywords{}

\maketitle

%\tableofcontents (a right mover)

\section{\label{sec:intro} Introduction}

The concept of `jet quenching' encompasses a variety of phenomena, among which
transverse momentum broadening, 
radiative energy loss, color decoherence, or medium--induced jet fragmentation,
which accompany the propagation of a `hard probe' 
(an energetic parton, or the jet generated by its evolution) through the dense QCD
medium created at the intermediate stages of an ultrarelativistic nucleus--nucleus
collision. % and which allow us to scrutinize the properties of this medium.
The theoretical description of these phenomena within perturbative QCD reveals a remarkable 
universality \cite{Baier:1996kr,Baier:1996sk,Zakharov:1996fv,Zakharov:1997uu,Baier:1998kq,Wiedemann:2000za,Wiedemann:2000tf,Arnold:2001ms,Arnold:2002ja,MehtarTani:2010ma,MehtarTani:2011tz,CasalderreySolana:2011rz,Blaizot:2012fh,Blaizot:2013hx,Blaizot:2013vha}:  to leading order in $\alpha_s=g^2/4\pi$ 
(the QCD coupling, assumed to be small), they all
depend upon the medium properties via a single quantity,  known as the `jet quenching parameter' 
$\hat q$. This quantity is a quasi--local transport coefficient which characterizes
the dispersion in transverse momentum accumulated by the fast parton after 
crossing the medium over a distance $L$: $\langle p_\perp^2\rangle \simeq\hat q L$.
The universality alluded to above holds because, to leading order in $\alpha_s$, there is
the same basic mechanism --- namely, in--medium collisions with a relatively large momentum
transfer and with a cross--section proportional to $\hat q$ --- which controls all the phenomena
associated with `jet quenching'.

Very recently, it has been shown \cite{Liou:2013qya,Iancu:2014kga,Blaizot:2014bha} 
that this universality persists after resuming the radiative corrections in the {\em
double--logarithmic approximation} (DLA),  that is, after taking into account a particular subset
of loop corrections where each power of $\alpha_s$ is enhanced by the double logarithm
$\ln^2(L/\lambda)$. Here, $L$ is the distance travelled by the `hard probe' through the medium and
$\lambda$ is the typical wavelength of a medium constituent (e.g., $\lambda=1/T$ for 
a weakly--coupled quark gluon plasma with temperature $T$). The DLA encompasses the dominant
radiative corrections in the limit of a large medium, $L\gg \lambda$. In particular, when
$\alpha_s\ln^2(L/\lambda)\gtrsim 1$, it becomes the leading--order approximation to the 
physics of jet quenching in pQCD.

The radiative corrections of interest for us here are associated with medium--induced emissions of 
soft gluons by the energetic parton. They naturally contribute to the parton transverse momentum
broadening $\langle p_\perp^2\rangle$, via their recoil, and also to the parton energy 
loss, via the energy taken away by the unresolved emissions. 
To DLA accuracy, all such effects %of the radiative corrections
can be simply taken into account via a renormalization of the jet quenching parameter $\hat q$. 
This is quite remarkable in several respects. First, the radiative corrections associated with bremsstrahlung
are generally non--local in time, due to the finite `formation time' for gluon emissions, and thus
could significantly alter the quasi--linear proportionality between $\langle p_\perp^2\rangle$ 
and the medium size $L$. Second, phenomena like $p_\perp$--broadening and the 
radiative energy loss {\em a priori} explore different aspects of the interactions
between the hard probe and the medium, and hence they could be differently affected by quantum
fluctuations. The reason why, at DLA, the sole effect of the quantum evolution is a renormalization of 
$\hat q$ is because the corresponding fluctuations are sufficiently mild: they have relatively soft energies,
and hence very short formation times $\tau \ll L$, and also relatively small transverse momenta 
$p_\perp^2\ll Q_s^2(L)$, with $Q_s^2(L)\equiv\hat q L$ the transverse resolution scale
relevant for the calculation of $p_\perp$--broadening. On the other hand, such fluctuations are 
still sufficiently hard --- in the sense that $p_\perp^2\gg Q_s^2(\tau)$,
with $Q_s^2(\tau)\equiv\hat q \tau$ the saturation momentum of the gluon
distribution in the medium --- to undergo only a single scattering during their formation.
Accordingly,  their effects can be absorbed into a renormalization of $\hat q$,
which thus becomes non--local (i.e., $L$--dependent), but only mildly.

The renormalization of $\hat q$ to the accuracy of interest is described by a relatively simple, 
linear, equation, which has been derived at fixed coupling 
\cite{Liou:2013qya,Iancu:2014kga,Blaizot:2014bha} and is shown in \eqn{qren} below. 
This equation describes the evolution of
the (renormalized) jet quenching parameter $\hat q(\tau,\bp^2)$ with the lifetime $\tau\le L$ and the
transverse momenta squared $\bp^2\ge Q_s^2(\tau)$ of the fluctuations. It is formally similar,
in the sense of involving the same splitting kernel, to the familiar `DLA equation' 
\cite{DeRujula:1974rf}  (a common limit of the DGLAP
\cite{Gribov:1972ri,Dokshitzer:1977sg,Altarelli:1977zs}  and BFKL 
\cite{Lipatov:1976zz, Kuraev:1977fs,Balitsky:1978ic} equations), but it differs from the
latter in an essential way, as we now explain (see Sect.~\ref{sec:evol}  for more details).
The standard DLA equation in the literature is a {\em genuinely linear} equation, which describes
gluon evolution via successive branching in the dilute regime; it applies e.g.
to jet fragmentation in the vacuum and to the gluon distribution in the proton
at not too small values of Bjorken's $x$. By contrast, the `DLA equation' of interest for us here
is a particular limit, valid to the desired accuracy, of a {\em genuinely non--linear}
evolution: that of the incoming parton and of the associated radiation, which undergo multiple 
scattering off the medium constituents. Alternatively, by boosting to a frame where the medium 
itself is moving fast while the `probe' is relatively slow, one can associate this evolution 
with the gluon distribution in the medium, and the non--linear effects with gluon saturation. 
As discussed in \cite{Iancu:2014kga}, this non--linear problem can be viewed as a
generalization of the BK--JIMWLK evolution \cite{Balitsky:1995ub,Kovchegov:1999yj,JalilianMarian:1997jx,JalilianMarian:1997dw,
Kovner:2000pt,Weigert:2000gi,Iancu:2000hn,Iancu:2001ad,Ferreiro:2001qy}
to the case of an extended target. (We recall that  the BK--JIMWLK equations describe the 
non--linear evolution towards saturation of the gluon distribution in a 
`shockwave' target, like the Lorentz--contracted nucleus in proton--nucleus collisions at high energy;
see e.g. \cite{Gelis:2010nm} for a review.)

The reason why this generally complicated evolution simplifies so drastically at DLA 
(and reduces to a linear equation for $\hat q$), 
is because the fluctuations which matter to this accuracy undergo only 
single scattering, as alluded to before. But as a matter of facts, the non--linear effects are 
still present in this equation, via the integration limits in \eqn{qren}, which act as a
`saturation boundary': they delimitate the phase--space for the single--scattering
approximation. This saturation boundary is specific to the problem at hand
(it is absent in the standard DLA equation) and is particularly important for the physics
of jet quenching. Indeed, as we shall review in the next section, the solution $\hat q(\tau,\bp^2)$ 
is needed in the vicinity of the `saturation line', i.e. for $\bp^2\sim Q_s^2(\tau)=\hat q\tau$,
where the phase--space restriction introduced by the saturation boundary is truly essential:
it qualitatively modifies the behavior of the function  $\hat{q}_s(\tau)\equiv
\hat q(\tau, \bp^2=Q_s^2(\tau))$,
including in the asymptotic regime at large evolution `time' $Y\equiv \ln(\tau/\lambda)$.
The difference w.r.t. the standard DLA solution in the literature 
\cite{DeRujula:1974rf,Kovchegov:2012mbw} 
becomes even more pronounced after including the effects
of the {\em QCD running coupling}, as we shall demonstrate in this paper.

%that we shall here address for the first time.

The experience with other evolution processes in perturbative QCD, either linear
(DGLAP, BFKL), or non--linear (BK-JIMWLK), demonstrates that the effects of the running 
of the coupling are quantitatively and even qualitatively important, including in the approach
towards saturation. As a general rule, their main effect is to considerably slow
down the evolution. For instance, the logarithm of the saturation scale $\ln Q_s^2(Y)$ 
for a `shockwave' target grows linearly with the rapidity $Y=\ln(1/x)$ at
fixed coupling, but only like $\sqrt{Y}$ after including the running of the coupling
\cite{Iancu:2002tr,Mueller:2002zm,Munier:2003sj}. To DLA accuracy, 
one may even argue --- by analogy with the corresponding discussion
for the DGLAP equation --- that the running of the coupling is truly a {\em leading--order} 
effect: the logarithmic dependence of the coupling upon the (transverse)
resolution scale modifies the systematics of the resummation for the transverse logarithms.
This motivates our present study of the consequences of a running coupling for \eqn{qren}.

As previously mentioned, the DLA equation \eqref{qren} has been established for fixed
coupling and its heuristic extrapolation to a running coupling (without an explicit calculation
of the respective loop corrections) is a priori ambiguous. Fortunately though the physical
origin of the various factors of $\alpha_s$ is quite clear, which makes it easy (if not rigorous)
to ascertain the respective scale dependences. There is first a global factor $\abar\equiv
\alpha_s N_c/\pi$ in the r.h.s. of \eqn{qren}, which originates from the vertex for soft and
quasi--collinear gluon emissions. The experience with the DGLAP equation instructs us to
evaluate this coupling at a scale set by the transverse momentum of the emitted gluon:
$\abar\to\abar(\bp_1^2)$. Furthermore, the initial condition for this equation, i.e.
the `tree--level' value $\hat{q}^{(0)}$ of the jet quenching parameter, is itself dependent
upon the QCD coupling. Its calculation to leading order is well understood 
\cite{Baier:1996kr,Baier:1996sk,Arnold:2001ms,Arnold:2002ja,Arnold:2008iy,Arnold:2008vd,CaronHuot:2008ni} and naturally 
leads to the choices for the running coupling exhibited in \eqn{qhat02}. Whereas
there might still be some ambiguity with these choices, this has 
no incidence on the physical results for $\hat q$, as we shall see.

Our strategy to solve the integral equation \eqref{qren}, or its running--coupling version 
\eqn{qren2rc} will consist
in performing successive iterations in which the integrals are 
analytically computed. This will allow us to express the respective
solutions in the form of truncated series, which are rapidly convergent.
In the case of a fixed coupling, we will be able to resum this series  and thus obtain exact analytic solutions for $\hat q(\tau,\bp^2)$, which generalize the respective 
result in Ref.~\cite{Liou:2013qya} to arbitrary transverse momenta $\bp^2\gtrsim Q_s^2(\tau)$
and to different initial conditions. In the case of a running coupling, we have not been able to resum
the iteration series in closed form. (Because of the saturation boundary, the structure of this
series turns out to be considerably more complicated than for the standard DLA equation with 
running coupling, whose exact solution is well known.) Yet, by fitting the behavior of this
series at large $Y\gg 1$, we shall numerically extract the first 
three terms in the asymptotic expansion of $\ln \hat{q}_s(Y)$, that is, all the terms which increase
with $Y$ (see \eqn{logqrc}).
%Here, $\hat{q}_s(Y)$ is the value of the function $\hat q(Y, \bp^2)$ for 
%$\bp^2=Q_s^2(Y)$ (`along the saturation line'),
%which is the quantity relevant for a study of jet quenching. 
Importantly, these three terms turn out to be {\em universal}, i.e. independent of the choice
of the initial condition $\hat{q}^{(0)}$, within the class of initial conditions to be
considered here. In particular, the dominant term grows like $\sqrt{Y}$ ; as expected,
this growth is slower
than the asymptotic behavior $\ln \hat{q}_s(Y)\propto Y$ observed at fixed coupling
\cite{Liou:2013qya,Iancu:2014kga,Blaizot:2014bha}.

Another remarkable feature of our result is that the asymptotic expansion 
for $\ln \hat{q}_s(Y)$ appears to be 
extremely similar to that of $\ln Q_s^2(Y)$ (the saturation momentum for a  
large nucleus, or `shockwave'), as computed in \cite{Mueller:2002zm,Munier:2003sj}.
This similarity refers not only to the $Y$--dependence of the first two terms in this
expansion (proportional to $\sqrt{Y}$ and respectively $Y^{1/6}$), but 
also to the respective numerical coefficients. We have no fundamental
explanation for this similarity, but in our opinion it
points out towards some universal (in the sense of target--independent) 
features in the high--energy evolution towards saturation. In particular, it suggests
a deep connection between the DLA equation for $\hat q$ and
the `BFKL equation with a saturation boundary' \cite{Mueller:2002zm} --- a linearized version
of the BK equation \cite{Balitsky:1995ub,Kovchegov:1999yj}, 
in which the non--linear effects are again implemented via boundary conditions, 
and which provides the framework for the calculations of $\ln Q_s^2(Y)$ in 
Refs.~\cite{Iancu:2002tr,Mueller:2002zm,Munier:2003sj,Triantafyllopoulos:2002nz}.

While conceptually interesting, the asymptotic behavior of $\hat{q}_s(Y)$
at large $Y$ is not necessarily relevant for the phenomenology of jet quenching. 
In Sect.~\ref{sec:fy}, we shall argue that the physically interesting values
for the phenomenology of heavy ion collisions 
at RHIC and the LHC are in the ballpark of $Y=2\div 3$. To characterize the pertinence
of the resummation for such values of $Y$, we shall numerically compute
the enhancement factor $\hat{q}_s(Y)/\hat{q}^{(0)}$. This leads to two interesting
conclusions. First, the asymptotic behavior is approached quite fast, namely for $Y\simeq 2$,
at least in so far
as the $Y$--dependence of $\ln \hat{q}_s(Y)$ is concerned. Second, for $Y\lesssim 4$,
the enhancement factor $\hat{q}_s(Y)/\hat{q}^{(0)}$ turns out to be roughly the same for both
fixed and running coupling, and for all sets of initial conditions. In particular, for $Y=3$
this factor is numerically close to 3, which represents a significant enhancement with potentially
important consequences for the phenomenology. We believe that this factor 3 is a robust prediction
of our present calculations.

\section{\label{sec:evol} The DLA evolution of the jet quenching parameter}

To start with, let us more precisely explain what we mean by the `renormalization equation for the
jet quenching parameter'. To be specific, we shall discuss this in the context of the transverse momentum 
broadening, but our conclusions are more general, as we shall later argue.
Consider an energetic quark  which crosses the medium over a distance $L$ and thus
acquires a transverse momentum\footnote{By `transverse' we mean the two
dimensional plane $\bx=(x^1,x^2)$ orthogonal to the parton direction of motion, conventionally
chosen along $x^3$.}  $\bp$ via rescattering in the medium. This transverse momentum can be either
directly transferred from the medium constituents via collisions, or represent the recoil associated
with unresolved gluon emissions, which are themselves triggered by the collisions in the medium.
The transverse--momentum distribution
$\rmd N/\rmd^2\bp$ of the outgoing quark can be computed
as the Fourier transform of the $S$--matrix for a quark--antiquark dipole which
propagates through the medium:
\begin{align}
 \label{ptbroad}
 \frac{\rmd N}{\rmd^2\bp} \,=\,
 \frac{1}{(2\pi)^2} \int_{\br} \rme^{-\rmi \bp \cdot \br} \,\mathcal{S}(\br),
%\langle \hat{S}_{\bx\by}  \rangle
\end{align}
where $\br$ is the transverse size of the dipole.
This `dipole' is merely a mathematical construction: its `quark leg'  
represents the physical quark in the direct amplitude, whereas 
the `antiquark leg' is the physical quark in the complex conjugate
amplitude. The dipole $S$--matrix $\mathcal{S}(\br)$ encodes the relevant 
information about the multiple scattering
between the incoming quark and the medium,
within the limits of the eikonal approximation.
Using $\mathcal{S}(0)=1$ (`color transparency'), one sees that
the distribution \eqref{ptbroad} is properly normalized: $\int \rmd^2\bp\, (\rmd N/\rmd^2\bp)=1$.

We assume that the medium is weakly coupled, such as a quark-gluon plasma (QGP) with a 
sufficiently high temperature $T$, or cold nuclear matter (CNM) with sufficiently
high density. To leading order in perturbative QCD, the dipole $S$--matrix
takes the following form  (see the next section for details) :
%can be parametrized in terms of the {\em jet quenching parameter} $\hat q$ :
\begin{align}\label{Sdip0}
 \mcal {S}^{(0)}(\br)\,\simeq\,\exp\left\{-\frac{1}{4}\,L \hat q^{(0)}(1/r^2)\,\br^2\right\}\,,
  %\qquad
  % \hat q(1/r^2)\,\equiv\,4\pi\alpha_s^2 C_F n_0 \ln\frac{1}{r^2\Lambda^2}\,.
  \end{align}
where the {\em tree--level jet quenching parameter} $\hat q^{(0)}(Q^2)$ is a slowly (logarithmically) 
varying function upon the transverse resolution
of the scattering process, as fixed by the dipole size: $Q^2=1/r^2$. 
For transverse momenta $p_\perp=|\bp|$ which are not too high  (see below), 
the integral in \eqn{ptbroad} is cut off by the function $\mathcal{S}^{(0)}(\br)$, which decreases
very fast at large $r$.
This leads us to introduce the medium {\em saturation momentum} $Q_s(L)$, via the condition
that, when $r=1/Q_s$, the exponent in \eqn{Sdip0}  becomes of $\order{1}$~:
\beq\label{Qs0}
 Q_s^2(L) \,=\,  L \hat q^{(0)}(Q_s^2)
\eeq
For transverse momenta $p_\perp\lesssim Q_s(L)$, the Fourier transform in \eqn{ptbroad} is controlled 
by dipole sizes $r\sim 1/Q_s$ and can be evaluated by replacing 
$\hat q^{(0)}(1/r^2)\simeq \hat q^{(0)}(Q_s^2)$ within the integrand\footnote{For
much larger momenta $p_\perp\gg Q_s(L)$,  the scale dependence of $\hat q^{(0)}(1/r^2)$ becomes
essential and leads to a spectrum ${\rmd N}/{\rmd^2\bp}$ which decreases as $1/p_\perp^4$ ;
 see e.g. the discussion in  Sect.~4.1 of Ref. \cite{Iancu:2014kga}.}. 
We thus find
 \begin{align}
 \label{ptGauss}
 \frac{\rmd N}{\rmd^2\bp} \,\simeq\,\frac{1}{\pi Q_s^2}\,\rme^{-{\bp^2}/{Q_s^2}}\,.\end{align}
This Gaussian distribution is the hallmark of a diffusive process --- a random walk in the
transverse momentum space, leading to a momentum broadening
 $\langle p_\perp^2\rangle\simeq Q_s^2$ ---, which is induced by a succession
of independent collisions in the medium. An important lesson from the above is that,
in order to study $p_\perp$--broadening, one needs the dipole $S$--matrix {\em in
the vicinity of the saturation line}, i.e. for dipole sizes  $r\sim 1/Q_s$. 
A similar conclusion applies for the calculation of the radiative energy loss via
the BDMPSZ  mechanism for 
medium--induced gluon radiation \cite{Baier:1996kr,Baier:1996sk,Zakharov:1996fv,Zakharov:1997uu,Baier:1998kq,Wiedemann:2000za,Wiedemann:2000tf,Arnold:2001ms,Arnold:2002ja}. 
%For definiteness, we shall use the language and the notations appropriate for a medium, but 
%make the contact with the corresponding shockwave problem, whenever appropriate.

Consider now the dominant quantum corrections to this tree--level picture. We shall assume that 
the incoming projectile (the quark in the above example) has a sufficiently high energy
$E\gtrsim \omega_c$, where $\omega_c(L)\equiv \hat q L^2/2$ is a relatively hard 
medium scale whose physical meaning should soon become clear, 
and that it crosses the medium over a sufficiently large distance $L\gg \lambda$,
with $\lambda$ the typical wavelength of the medium constituents (e.g. $\lambda=1/T$ for a QGP
and $\lambda=1/m_N$, with $m_N$ the nucleon mass, for CNM).
%More precisely, we work under the assumptions that $\omega_c\gg Q_s(L)\gg T$.
Then, as shown in Refs.~\cite{Liou:2013qya,Iancu:2014kga,Blaizot:2014bha}, the interactions between
the projectile and the medium receive large radiative corrections, enhanced by the double 
logarithm\footnote{See also Ref.~\cite{Wu:2011kc} for a similar but earlier observation, which 
has motivated the more elaborate analysis in Ref.~\cite{Liou:2013qya}.} 
$\ln^2(L/\lambda)$. For instance, the transverse momentum broadening receives one--loop corrections
that can be cast into the form $\delta \langle p_\perp^2\rangle=\delta\hat q(L) L$, with
$\delta\hat q(L)\sim \abar\ln^2(L/\lambda)$ and $\abar\equiv\alpha_sN_c/\pi$ \cite{Liou:2013qya}.

\begin{figure}
\begin{center}
\includegraphics[scale=0.33]{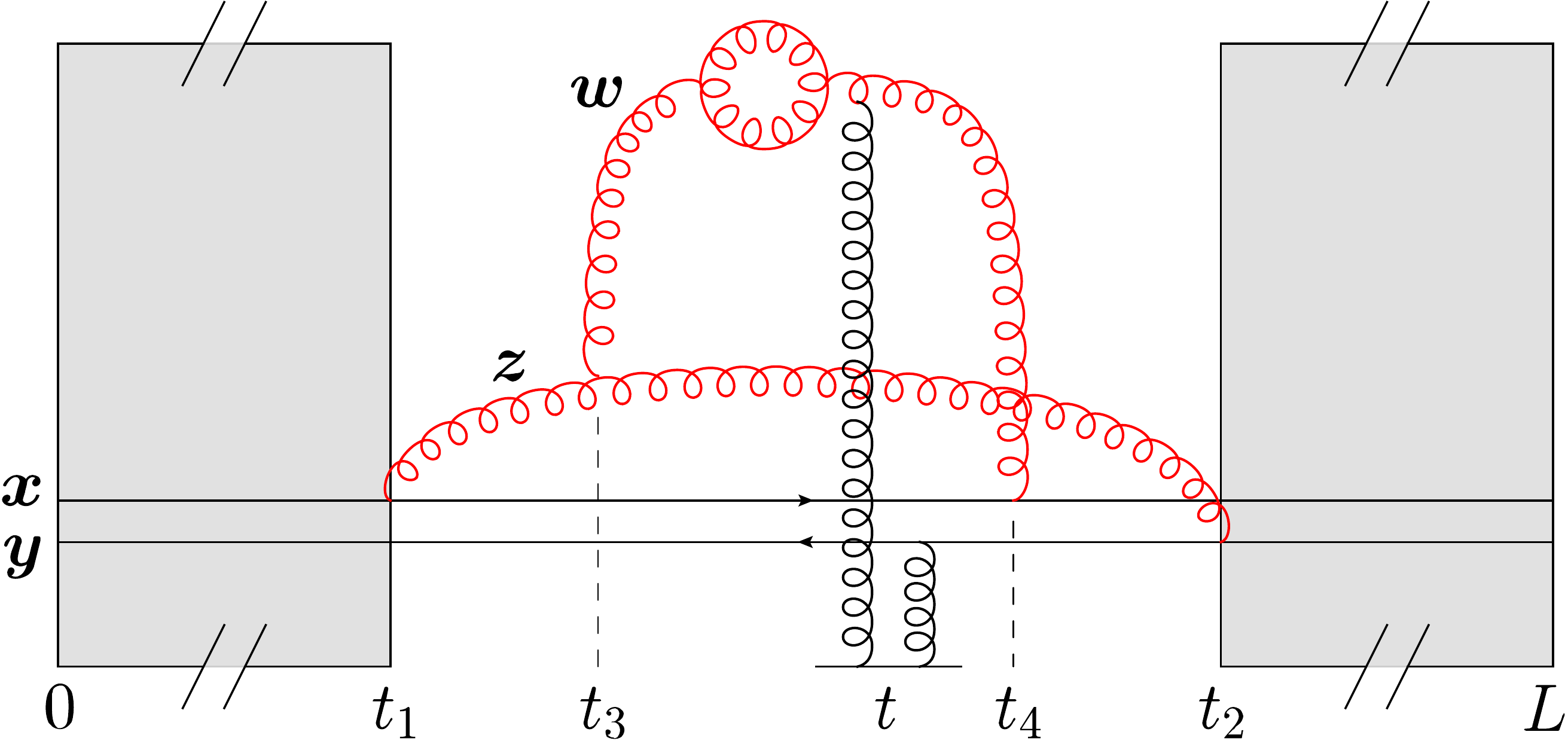}
\end{center}
\caption{\sl A Feynman graph representative for the evolution of jet quenching  in the double logarithmic approximation and with a running coupling. This graphs exhibits a parton
cascade made with 2 gluons successively radiated in the evolution (the gluon at $\bz$ being
radiated before the one at $\bw$), together with a gluon loop representative of the running coupling
effects. The whole cascade suffers a single scattering, at some
intermediate time $t$. As suggested by this picture, successive gluons in the evolution
have shorter and shorter lifetimes and larger and larger transverse sizes (which are all
much larger than the size $r=|\bx-\by|$ of the original dipole).}
\label{fig:evol}
\end{figure}

These corrections are associated with gluon fluctuations (unresolved gluon emissions with 
energies $\omega<\omega_c$) which in the plasma rest frame are naturally interpreted as 
bremsstrahlung by the projectile: these emissions are soft relative to their parent quark
(typically, $\omega\ll\omega_c< E$), but hard as compared to the medium
constituents ($\omega \gg 1/\lambda$). The double--logarithmic enhancement comes
from integrating over the (energy and transverse) phase--space for these emissions and, more
precisely, over the particular domain in this phase--space which corresponds to fluctuations which
{\em scatter only once} during their formation time $\tau=2\omega/p_\perp^2$. Specifically, one logarithm
is generated by integrating over $\tau$ within the range $\lambda \ll \tau \ll L$, and the other by
integrating over the gluon transverse momentum $p_\perp$, within the range
$\hat q\tau \ll p_\perp^2 \ll Q_s^2(L)$. The lower limit $\hat q\tau$ on $p_\perp^2$ is the condition
of single scattering and will play a major role in what follows. 
%This can be also interpreted in terms of gluon saturation in the medium, as we shall later discuss.
Gluon emissions with lower momenta $p_\perp^2\lesssim \hat q\tau$ are possible as well,
but they undergo multiple scattering and do not contribute to double--logarithmic accuracy.
(But they do contribute to single--logarithmic accuracy: they generate corrections of 
order $\abar\ln(L/\lambda)$ \cite{Liou:2013qya,Iancu:2014kga}.) Notice that gluons with energy $\omega
\sim \omega_c(L)$ and transverse momentum $p_\perp\sim Q_s(L)$ have a formation time $\tau= L$~; 
hence, $\omega_c$ is the upper limit on the energy of the medium--induced emissions 
\cite{Baier:1996kr,Baier:1996sk,Zakharov:1996fv}.

These large one--loop corrections represent only the first step in a quantum evolution, which can
be viewed as an evolution with increasing the `medium size' $L$ (more precisely, the distance $L$
travelled by the projectile through the medium) for a given value of $\lambda$. Namely, 
the primary gluons emitted by the projectile can in
turn radiate even softer gluons, thus giving rise to gluon cascades, which
are enhanced by the phase--space: the powers of $\abar\equiv\alpha_sN_c/\pi$ associated 
with soft gluon emissions can be accompanied by either double, or at least single, 
logarithms of $L/\lambda$, depending upon the kinematics of the emissions.
This evolution is generally {\em non--linear}, because of the multiple 
scattering between the partons in the cascades and the medium. Alternatively, the non--linear effects
can be viewed as {\em gluon saturation} in the medium \cite{Iancu:2014kga} (see also below). 

A theoretical framework which encompasses this evolution to leading--logarithmic accuracy
has been recently developed in Ref.~\cite{Iancu:2014kga}. The general evolution equations
appear to be extremely complicated, because of the non--locality in time of the multiple scattering
(the emitted gluons can scatter all the way during their propagation through the medium) and,
related to that, because of the failure of the eikonal approximation for the evolution gluons. In particular,
the dipole--$S$ matrix obeys a non--linear equation which can be viewed as a functional and
non--local generalization of the BK equation \cite{Balitsky:1995ub,Kovchegov:1999yj}.
This equation is probably too complicated to be solved in general. 
Fortunately though, the double--logarithmic corrections, 
which are the {\em dominant} corrections in the limit where $L/\lambda\gg 1$, are comparatively 
simple and easy to extract from the general evolution --- precisely because they are associated with
single scattering alone. These 
corrections form an `island' of (effectively) {\em linear} evolution, where the non--linear effects enter
only via the boundaries of the phase--space (namely, via the condition $p_\perp^2\gg\hat q\tau$ alluded 
to above). A Feynman graph representative for these corrections is illustrated in Fig.~\ref{fig:evol}.

As shown in \cite{Iancu:2014kga}, the linear evolution in the {\em double logarithmic approximation} (DLA)
preserves the same functional form for  $\mathcal{S}(\br)$ as at tree--level, cf. \eqn{Sdip0},
except for the replacement of the tree--level jet quenching parameter $\hat q^{(0)}$ by
a renormalized `value' (actually, a function of two variables; see below) $\hat q$~:
 \begin{align}\label{SDLA}
 \mcal {S}(L,\br)\,\simeq\,\exp\left\{-\frac{1}{4}L \hat q(L,1/r^2)\,\br^2\right\}\,.
 \end{align}
In turn, this implies the {\em universality} of the evolution to this accuracy:
all the quantities that can be computed from the dipole $S$--matrix ($p_\perp$--broadening,
radiative energy loss, BDMPSZ spectrum) get renormalized simply via
the appropriate redefinition of $\hat q$. 

\eqn{SDLA} involves the function $\hat q(\tau,\bp^2)$ which represents the renormalized jet quenching 
parameter as obtained after integrating our fluctuations with lifetimes up to $\tau$ and transverse
momenta up to $\bp^2$, the double--logarithmic accuracy. This function has support at  
$\bp^2\ge \hat q\tau$, where it is defined as the solution to the following integral equation
\cite{Liou:2013qya,Iancu:2014kga,Blaizot:2014bha} :
 \beq
 \label{qren}
 \hat{q}(\tau,\bp^2) = 
 \hat{q}^{(0)} +
 \abar \int_{\lambda}^{\tau} 
 \frac{\dif \tau_1}{\tau_1}
 \int_{\hat{q} \tau_1}^{\bp^2}
 \frac{\dif \bp_1^2}{\bp_1^2}\,
 \hat{q}(\tau_1,\bp_1^2)\,,
 \eeq 
to be subsequently referred to as the {\em DLA equation with a saturation boundary}. 
The  `saturation boundary' is the lower
limit $\hat{q} \tau_1$ in the above integral over $\bp_1^2$, which expresses the single--scattering condition, 
as already mentioned. This condition can be also understood with reference to gluon saturation in the
medium \cite{Iancu:2014kga}: the quantity $x\equiv \lambda/\tau$ represents the longitudinal
momentum fraction of the gluons from the medium which participate in the scattering\footnote{This 
interpretation holds in a Lorentz frame where the projectile is relatively slow, 
whereas the medium is highly boosted. In that frame, the gluon fluctuations involved in the 
evolution belong to the gluon distribution in the medium. That is, they
are Weizs\"acker--Williams quanta emitted by the
medium constituents; see \cite{Iancu:2014kga} for details.}, whereas
$ \hat q\tau=\hat q\lambda/x\equiv
Q_s^2(x)$ is the plasma saturation momentum for a given $x$.
(The `global' scale $Q_s^2(L)$ introduced before is the maximal possible value of $Q_s^2(x)$,
corresponding to $x=x_{\rm min} \equiv \lambda/L$.) 
Then the condition $p_\perp^2\gg\hat q\tau$ can be rewritten in the more familiar
form $p_\perp^2 \gg Q_s^2(x)$, 
which defines the dilute part of the gluon distribution in the medium.
Because of the presence of this saturation boundary, \eqn{qren} differs from the
more familiar `DLA equation' encountered in studies of the BFKL or DGLAP evolutions,
and this difference has profound physical consequences, as we shall see.
%In fact, \eqn{qren} 

For our present purposes, it is more convenient to view the saturation momentum as a function
of the lifetime $\tau$ of the gluon fluctuations, rather than of their longitudinal momentum fraction $x$.
Given the solution $\hat q(\tau,\bp^2)$ to \eqn{qren}, the  saturation momentum $Q_{s}^2(\tau) $
is implicitly defined by the following equation, which generalizes \eqn{Qs0} :
 \beq\label{Qsx}
 Q_{s}^2(\tau) \,=\,\tau\,\hat q\big(\tau, \bp^2=Q_{s}^2(\tau)\big)\,.\eeq
The value of the function $Q_{s}^2(\tau)$ at the `physical point' $\tau=L$ determines the 
$p_\perp$--spectrum according to \eqn{ptGauss}.
But in order to compute the function $\hat{q}(\tau,\bp^2)$ along the saturation line
$\bp^2=Q_{s}^2(\tau)$, and in particular at $\tau=L$,
we need to first solve the non--local equation \eqref{qren} for generic values
$\tau$ and $\bp^2 \ge \hat q\tau$. 

Note also that the lower integration limit $\hat{q} \tau_1$ in \eqn{qren} should be more
precisely understood as $Q_{s}^2(\tau_1)$. (The notation ``$\hat{q} \tau_1$'' becomes ambiguous
when $\hat{q}$ is a non--trivial function of $\tau$ and $\bp^2$.) Since the  saturation momentum 
$Q_{s}^2(\tau) $ is itself determined by the solution to \eqn{qren}, cf. \eqn{Qsx}, 
one may wonder what should be the respective value to be used in the integration limit in \eqn{qren}.
We shall later show that, to the desired accuracy, this can be safely taken as the {\em initial}
saturation momentum, as determined by \eqn{Qs0}. Furthermore, still
for that purpose, one can neglect the mild dependence of $\hat{q}^{(0)}$ upon the transverse momentum
$\bp^2$, that is, one can treat $\hat{q}^{(0)}$ as a constant within the integration limit.
%Accordingly, from now on, we shall write the `saturation boundary' in \eqn{qren}  simply
%as $\hat{q}^{(0)}\tau_1$. 
A similar approximation is authorized in the calculation of the renormalized saturation momentum
according to \eqn{Qsx} : to DLA accuracy, one can write
$Q_{s}^2(\tau)=\tau\hat{q}_s(\tau)$ with $\hat{q}_s(\tau)\equiv\hat q(\tau,\bp^2=\hat q^{(0)}\tau)$.
 
 For what follows, it is useful to rewrite \eqn{qren} in terms 
of the logarithmic variables $Y\equiv\ln(\tau/\lambda)$ and $\rho\equiv\ln (\bp^2/\hat{q}^{(0)}\lambda)$. 
 For arbitrary $Y$ and $\rho$ with $\rho \ge Y$, we have
 \beq
 \label{qren2}
 \hat{q}(Y,\rho) = \hat{q}^{(0)}+\abar
 \int_{0}^{Y} \dif Y_1 
 \int_{Y_1}^{\rho} \dif \rho_1\,
 \hat{q}(Y_1,\rho_1).
 \eeq 
Note that the tree--level  saturation line $\bp^2=\hat q^{(0)}\tau$ corresponds to $\rho=Y$. After
evolution, this becomes $Q_s^2(Y) =\lambda \rme^{Y}\hat q_s(Y)$, with $\hat q_s(Y)=
\hat{q}(Y,\rho=Y) $.

One obvious question refers to the validity limits of the present approximation. \eqn{qren}
or \eqref{qren2} resums the double--logarithmic corrections, but not also the single--logarithmic ones.
Hence, clearly, \texttt{(i)} this resummation becomes necessary when $L/\lambda$ is large enough for
$\abar\ln^2(L/\lambda)\gtrsim 1$ (or $\abar Y^2 \gtrsim 1$),  \texttt{(ii)} it correctly provides the dominant
asymptotic behavior in the large--medium limit $\abar\ln^2(L/\lambda)\gg 1$ (or $\abar Y^2 \gg 1$), 
and \texttt{(iii)} it has an intrinsic error of relative order $\abar\ln(L/\lambda)$ (or $\abar Y$).
This implies that the DLA can be strictly trusted for $Y$--values within a window
\beq\label{rangeY}
\frac{1}{\sqrt{\abar}}\ \lesssim \ Y \ < \ \frac{1}{\abar}\,,\eeq
which is parametrically large when $\abar \ll 1$, but in practice is admittedly quite limited.
Yet, as alluded to above, the dominant asymptotic behavior of the DLA solution can be
trusted for arbitrarily large $Y$. This conclusion will be substantiated by the subsequent analysis 
of \eqn{qren2}, which will allow us to more precisely characterize the accuracy of the DLA
in relation with the asymptotic behavior.

So far, our whole discussion has been carried at strict leading order in pQCD, meaning 
that the coupling $\abar$  in \eqn{qren2} is {\em a priori} fixed. But as well known from the experience
with pQCD evolution, the inclusion of running coupling effects is truly essential in order 
to obtain realistic estimates (in particular, for applications to the phenomenology).
This is particularly important for the problem at hand, in view of the strong non--locality of 
\eqn{qren2} in the transverse phase--space, which is logarithmic. The situation is reminiscent 
in that respect of the familiar DGLAP equation \cite{Gribov:1972ri,Dokshitzer:1977sg,Altarelli:1977zs} 
and its DLA limit \cite{DeRujula:1974rf} : there is no fundamental distinction between the 
transverse logarithms coming from the integration over the phase--space and those introduced 
by the running of the coupling. In that sense, the running coupling effects count already to leading
order. To cope with that, we shall introduce the one--loop running of the coupling according to
 \beq
 \label{arun}
 \abar(\rho) = \frac{b }{\ln (\bp^2/\Lambda^2)} = 
 \frac{b }{\rho + \rho_0}
 \quad \mbox{with} \quad
 b  = \frac{12 N_c}{11 N_c - 2 N_f}\,.
 \eeq
Whenever we shall need a specific value, we shall chose 
$N_f=N_c=3$, and hence $b =4/3$. The `constant shift' 
$\rho_0 = \ln\big(\hat{q}^{(0)} \lambda/\Lambda^2\big)$ 
in \eqn{qren2rc} emerges 
from the fact that, even though $\rho$ is logarithmically related to $\bp^2$, the reference scale 
is not $\Lambda^2$.
As we shall later see, the  presence  of this reference scale $\rho_0$ 
is truly important for numerical estimates, in particular, for
the phenomenology. 

Returning to \eqn{qren2}, it is quite clear (especially in view of the experience with the
DGLAP evolution) that the proper scale for evaluating the factor $\abar$ is the `running' 
scale $\rho_1$. This brings us to the following form for
the DLA evolution equation with running coupling (RC):
 \beq
 \label{qren2rc}
 \hat{q}(Y,\rho) = \hat{q}^{(0)}+b 
 \int_{0}^{Y} \dif Y_1 
 \int_{Y_1}^{\rho} \frac{\dif \rho_1}{\rho_1+\rho_0}\,
 \hat{q}(Y_1,\rho_1).
 \eeq
 
% \comment
{
 
\section{\label{sec:ic} The jet quenching parameter at tree-level}

In order to solve the DLA equations \eqref{qren2} and \eqref{qren2rc} via iterations, one needs
more information about the initial condition $\hat{q}^{(0)}$, namely, one needs to know its dependence
upon the resolution scale $\rho$ and also (in view of the RC problem) upon the QCD
coupling. In this section, we shall briefly recall the leading order calculation of $\hat{q}^{(0)}$, with
emphasis on the two aspects alluded to above. We refer to the literature for more detailed discussions
\cite{Baier:1996kr,Baier:1996sk,Arnold:2001ms,Arnold:2002ja,Arnold:2008iy,Arnold:2008vd,CaronHuot:2008ni}.
At leading order, the argument of the running coupling is ambiguous and will be fixed from
physical considerations. To that aim, we need to carefully
keep trace of the physical origin of the various 
factors of $\alpha_s$.

As explained in the previous section, our starting point is the dipole $S$--matrix 
$\mathcal{S}(\br)$, that we shall here evaluate at tree--level. We assume that the energy of the 
dipole, as measured in the rest frame of the medium, is high enough for the eikonal approximation
to be applicable. In this approximation, the effect
of the interaction is a color precession of the quark and the antiquark by the (fluctuating) color
field representing the (gluon distribution of the) medium. Assuming the dipole to be a right--mover,
and hence the medium to be a left--mover, the dipole $S$--matrix is computed as
\begin{align}\label{Sdip} \mcal{S}(\bx,\by)\,=\,  \frac{1}{N_c} \,\Big\langle
\rmtr\big[ V^{\dagger}({\bx}){V}({\by}) \big]\Big\rangle\,,\qquad
V^{\dagger}(\bx) = \rmP \exp\left\{\rmi g \int \dif x^+ \,A^-_a(x^+, {\bx}) t^a\right\}\,.
\end{align}
Here $\bx$ and $\by$ are the transverse coordinates of the quark, respectively, the antiquark,
which are not changed by the interaction, $V^{\dagger}({\bx})$ and ${V}({\by})$ are Wilson lines
describing the respective color precessions, $A^-_a(x^+, {\bx})$ is the `large' component of the
color field in the target, which is randomly distributed (due to quantum and thermal fluctuations),
and the brackets $\langle\cdots\rangle$ refer to the average over this `background' field.
The light--cone coordinate $x^+\equiv (x^0+x^3)/\sqrt{2}$ plays the role of a `time' 
for the dipole and, respectively, of a longitudinal coordinate for the medium. The $t^a$'s are 
the color group generators in the fundamental representation
and P stands for path ordering w.r.t. $x^+$. 

%$\br\equiv \bx - \by$

A weakly--coupled medium can be described as a collection of independent
color charges --- thermal quarks and gluons for a QGP with sufficiently high temperature $T$,
or valence quarks for dense CNM, as described in the McLerran--Venugopalan (MV) model 
\cite{McLerran:1993ni,McLerran:1994vd}. These charges will be assumed to be
point--like and have no other mutual interactions,
except for those responsible for the screening of the color interactions over a transverse
distance $r\sim 1/\Lambda$. For a weakly--coupled QGP, this screening is perturbative
and $\Lambda=m_D$, with $m_D^2\sim \abar T^2$ 
the Debye mass. Here, however, we shall mostly focus on the case of CNM, where the
screening is non--perturbative and associated with confinement. 
Under these assumptions, the only non--trivial correlator of the target 
field $A^-$ is the respective 2--point function, which has the following structure
 \begin{align}\label{correlmed}
\left\langle A^-_a(x^+,x^-,\bx)\,A^{-}_b(y^+,y^-,\by)\right\rangle\,=\,\delta_{ab}\delta(x^+-y^+)\,n_0
\gamma(\bx-\by)\,,
\end{align}
where $n_0$ is the number  density of the medium constituents,
weighted with appropriate color factors. For simplicity we assume the medium to be
homogeneous.  Also
  \begin{align}\label{Coul}
 \gamma(\bk)\equiv \int \rmd^2\br\ \rme^{\rmi \bk\cdot\br}\,\gamma(\br)\,\simeq\,
 \frac{g^2}{\bk^4}\,,\end{align}
(with the approximate equality holding for $k_\perp\gg \Lambda$)
is the square of the 2--dimensional Coulomb propagator. It is understood
that \eqn{Coul} must be used with an infrared cutoff $k_\perp\simeq \Lambda$. 

For the Gaussian field distribution in \eqn{correlmed}, it is a straightforward exercise to
compute the average $S$--matrix for a quark--antiquark dipole.  One finds (with $\br\equiv \bx - \by$)
  \begin{align}\label{Sdip0}
 \mcal{S}^{(0)}(\br)\,=\,\exp\Biggl\{-g^2 C_F n_0 L
 \int \frac{\rmd^2\bm{k}}{(2\pi)^2}\,\gamma(\bk)\left(1-
 \rme^{\rmi\bm{k}\,\cdot\,\bm{r}}\right)\Biggr\}\,.\end{align}
Using \eqn{Coul}, one sees that the integral over $\bk$ in \eqn{Sdip0} is logarithmically
sensitive to the IR cutoff $\Lambda$. We shall be mostly interested in
small dipole sizes $r\equiv |\br|\ll 1/\Lambda$. 
Then, there is a large  logarithmic phase--space, at $\Lambda\ll k_\perp\ll 1/r$. 
To leading logarithmic accuracy, the integral can be evaluated by expanding the complex
exponential $\rme^{\rmi\bm{k}\,\cdot\,\bm{r}}$ to second order (the linear term vanishes
after angular integration). One thus finds the result previously shown in \eqn{Sdip0},
with the following expression for $\hat q^{(0)}(Q^2)$ (we recall that $Q^2=1/r^2\gg \Lambda^2$) :
   \begin{align}\label{qhat}
   \hat q^{(0)}(Q^2) \,\simeq\,4\pi\alpha_s(Q^2) C_F n_0\int^{Q^2}_{\Lambda^2}
    \frac{\rmd\bm{k}^2}{\bk^2}\, \alpha_s(\bk^2) \,\sim\,\alpha_s(Q^2) \,n_0
    \,xG(x,Q^2)\,.\end{align}
This is the tree--level jet quenching parameter
 for an incoming quark. The corresponding quantity for a gluon is obtained by multiplying 
\eqn{qhat} with $N_c/C_F$. As also shown above, this expression is usually written as a
proportionality between $\hat q^{(0)}(Q^2)$ and the gluon distribution $xG(x,Q^2)$
produced by one parton in the medium, on the resolution scale $Q^2$.

In writing \eqn{qhat}, we have already ascribed the running scales for each of the two
factors of $\alpha_s$, on physical grounds: \texttt{(i)} The factor $\alpha_s$ outside the integral 
originates in the coupling between the (anti)quark in the dipole and the target color field, 
via the Wilson lines in \eqn{Sdip}; the transverse resolution for this interaction is fixed by the dipole
size, $Q^2=1/r^2$, so this is the natural argument for the running of that coupling.
\texttt{(ii)} The factor $\alpha_s$ inside the integral comes from the Coulomb propagator \eqref{Coul},
so it naturally runs with the momentum $k_\perp^2$ exchanged via Coulomb scattering.

In the case of a fixed coupling,  the integral over $\bk^2$ generates a logarithmic
enhancement:
  \begin{align}\label{q0FC}
 \hat q^{(0)}(Q^2) \,\simeq\,
 4\pi\alpha_s^2 C_F n_0  \ln\frac{Q^2}{\Lambda^2}\qquad\mbox{(fixed coupling)}\,.\end{align}
On the other hand, with a running coupling, the integral over $\bk^2$ yields only
 a mild enhancement:
  \begin{align}\label{q0RC}
 \hat q^{(0)}(Q^2) \,\simeq\,
 4\pi\alpha_s(Q^2) C_F n_0  b_0 \ln {\ln ({Q^2}/{\Lambda^2})}
 %\, \simeq\,4\pi b_0^2 C_F n_0 \frac{\ln(\rho+\rho_0)}{\rho+\rho_0}
 \qquad\mbox{(running coupling)}\,,\end{align}
  where $b_0$ is related to the parameter $b$ introduced in \eqn{arun} via $b=(N_c/\pi)b_0$.

 One can summarize the above results as follows:
  \beq
 \label{qhat02}
 \hat{q}^{(0)}(\rho) =
 \begin{cases}
 \tilde{q}^{(0)}(\rho+\rho_0)
 &\mbox{for fixed coupling},
 \\*[0.2cm]
 \tilde{q}^{(0)}\displaystyle{\frac{\ln (\rho+\rho_0)}{\rho+\rho_0}}
 &\mbox{for running coupling}.
 \end{cases}
 \eeq
 where\footnote{Strictly speaking, one should write $\rho= \ln (Q^2/Q_0^2)$ and
 similarly $\rho_0= \ln (Q_0^2/\Lambda^2)$, with $Q_0^2$ the minimal saturation momentum
 at tree--level, defined by $Q_{0}^2 = \hat{q}^{(0)}(Q_0^2)\lambda$. Since, however,
 the scale dependence of the tree--level jet quenching 
 parameter is very mild (for either fixed or running coupling), we can replace
 $Q_{0}^2 \to \hat{q}^{(0)}\lambda$, where $\hat{q}^{(0)}$ is effectively treated as a constant.}
  $\rho= \ln (Q^2/\hat{q}^{(0)}\lambda)$, $\rho_0= \ln (\hat{q}^{(0)}\lambda/\Lambda^2)$,
 and it is understood that the $\rho$--independent prefactor $ \tilde{q}^{(0)}$ is different for
 fixed and respectively running coupling, although we shall use the same notation in both cases.

In what follows, we shall study the DLA solutions with the initial conditions given by \eqn{qhat02}, 
and also those where $\hat{q}^{(0)}$ is taken to be simply a {\em constant}. The latter
choice allows for simpler analytic manipulations, while yielding the same asymptotic
behavior at large $Y=\rho$ --- for both
fixed and running coupling --- as the more `realistic' initial conditions in \eqn{qhat02}.
This points out towards the universality of the large--$Y$ asymptotics w.r.t. the choice of the 
initial conditions.

\section{\label{sec:fc} The exact solution for fixed coupling}

It turns out that it is rather straightforward to solve
the fixed coupling evolution equation \eqref{qren2} via successive iterations.
For a constant (i.e. $\rho$--independent) initial condition and on the (tree--level)
saturation line $\rho=Y$, the corresponding solution --- to be denoted here as $\hat q_s(Y)$ ---
has already been constructed in this in Ref.~\cite{Liou:2013qya}. In this section we shall
extend the solution in  \cite{Liou:2013qya} to generic values of $\rho$ and also to the more
realistic,  $\rho$--dependent,  initial condition shown in the first line of \eqn{qhat02}.

For reasons to shortly become clear, it is convenient to perform the $\rho_1$--integration in \eqn{qren2}  
in the full available space, that is from 0 to $\rho$, and then subtract the contribution which is cut by the saturation boundary. That is, we shall write one iteration step as
 \beq
 \label{qiter}
 \hat{q}^{(n)}(Y,\rho) =
 \abar
 \int_{0}^{Y} \dif Y_1 
 \int_{0}^{\rho} \dif \rho_1\,
 \hat{q}^{(n-1)}(Y_1,\rho_1)
 -\abar
 \int_{0}^{Y} \dif Y_1 
 \int_{0}^{Y_1} \dif \rho_1\,
 \hat{q}^{(n-1)}(Y_1,\rho_1),
 \eeq 
and then the final solution will be given by the summation of the series
 \beq
 \label{qseries}
 \hat{q}(Y,\rho) =
 \sum_{n=0}^{\infty}\hat{q}^{(n)}(Y,\rho). 
 \eeq
Our notation is such that $\hat{q}^{(n)}$ is the correction of order $\abar^n$. Assuming a momentum independent initial condition $\hat{q}^{(0)}$, it is trivial to do the first iteration and find
 \beq
 \label{q1}
 \hat{q}^{(1)}(Y,\rho)=
 \hat{q}^{(0)} \abar Y \rho
 -\hat{q}^{(0)} \abar\, \frac{Y^2}{2},  
 \eeq
where the two terms correspond to the two terms in \eqn{qiter}. Proceeding to the second iteration one may naively expect that there will be four terms, however one finds the that two terms generated by inserting \eqn{q1} into the `subtracted' contribution in \eqn{qiter} exactly cancel each other. Remarkably,
this pattern repeats itself to all subsequent orders in $\abar^n$ ;
that is, only the first term in \eqn{qiter} contributes to the calculation 
of $\hat{q}^{(n)}(Y,\rho)$ for $n\geq2$. It is then straightforward to deduce that for a generic $n$ one has
 \beq
 \label{qn}
 \hat{q}^{(n)}(Y,\rho)=
 \hat{q}^{(0)}\, \frac{(\abar Y \rho)^n}{(n!)^2}
 -\hat{q}^{(0)}(1-\delta_{n0})\, \frac{Y}{\rho}\,
 \frac{(\abar Y \rho)^n}{(n-1)!(n+1)!},
 \eeq  
where the factor $1-\delta_{n0}$ removes the second term when $n=0$. 
The two terms above are easily recognized as belonging to the Taylor expansions of two
modified Bessel functions of the first kind, $\rmI_0$ and respectively $\rmI_2$~:
 \beq
 \label{qyrho}
 \hat{q}(Y,\rho) =
 \hat{q}^{(0)}
 \bigg[
 \rmI_0\big(2 \sqrt{\abar Y \rho}\big) 
 - \frac{Y}{\rho}\,\rmI_2\big(2 \sqrt{\abar Y \rho}\big)
 \bigg].
 \eeq
 By itself, the first term alone would be the solution to the standard DLA equation, which has no saturation
 boundary \cite{DeRujula:1974rf,Kovchegov:2012mbw}; that is, this is the solution that would be obtained
 by iterating the first term in the r.h.s. of \eqn{qiter} alone.
 Vice--versa, the second, negative, term in \eqn{qyrho} is entirely
 due to the presence of the saturation boundary and is reminiscent of a corresponding term emerging 
 in the calculation\footnote{Cf.~Eq.~(40) in \cite{Mueller:2002zm}, where one cuts the contributions coming from $Q \lesssim Q_s$ by subtracting a term of similar structure.}
  of the saturation momentum for a `shockwave' (a Lorentz contracted, large, nucleus)  
 \cite{Mueller:2002zm}. In that case too, the evolution of the dipole $S$--matrix in the vicinity
 of the saturation line can be approximately described by a linear equation with a saturation boundary,
 but the respective linear equation is BFKL \cite{Lipatov:1976zz, Kuraev:1977fs,Balitsky:1978ic}
 (the linearized version of BK equation), and not DLA.
  
The asymptotic expansion of the modified Bessel function $\rmI_n(x)$ for large values of
its argument $x$ reads 
 \beq
 \label{besselasym}
 \rmI_n(x) = \frac{\rme^x}{\sqrt{2\pi x}}
 \bigg(1+\frac{1-4 n^2}{8x} + \cdots \bigg).
 \eeq
 That is, both the leading exponential behavior and the leading term in the prefactor are
 independent of $n$.
This implies that the dominant exponential behavior of the solution 
\eqref{qyrho} for large values of $x=2 \sqrt{\abar Y \rho}$  is 
the same as it would be in the absence of the boundary. Most likely, this feature has no
fundamental meaning since, as we shall later discover, it is in fact washed out by the running of
the  coupling.

When evaluating \eqn{qyrho} on the (tree--level) saturation line at $\rho=Y$, the leading order term in the asymptotic
expansion at large $x=2 \sqrt{\abar}\, Y$ (the unity within the brackets in \eqn{besselasym}) precisely
cancels between the two terms in \eqn{qyrho}. Accordingly, the suppression due to the boundary manifests itself
at large $Y$ as an additional $1/Y$ prefactor.  Again, this is similar to the corresponding problem for a 
shockwave, as controlled by the BFKL equation with a saturation boundary: in that case too, the 
boundary introduces an extra prefactor $1/Y$ in the dominant behavior of the dipole amplitude
at large $Y$ and in the vicinity of the saturation line  \cite{Mueller:2002zm}.

As a matter of facts, for $\rho=Y$ one can combine the two terms in \eqn{qyrho} to get
a rather compact expression (the second equality below holds for $Y \gg 1/\sqrt{\abar}$),
 \beq
 \label{qyy}
 \hat{q}_s(Y) = \hat{q}^{(0)}\,
 \frac{\rmI_1\big(2 \sqrt{\abar}\, Y\big)}{\sqrt{\abar}\,Y}
 = \hat{q}^{(0)}\, 
 \frac{\rme^{2 \sqrt{\abar}\, Y}}{\sqrt{4\pi}\,(\sqrt{\abar}Y)^{3/2}}
 \left[1 + \mcal{O}(1/\sqrt{\abar}Y) \right],
 \eeq    
which agrees with the corresponding result in Ref.~\cite{Liou:2013qya}. The additional prefactor
$1/\sqrt{\abar}\,  Y$ is manifest on \eqn{qyy}. The renormalized saturation 
momentum is then obtained as %(cf. \eqn{Qsx})
\beq\label{QsYFC}
Q_s^2(Y)=\lambda \hat{q}_s(Y) \rme^Y \simeq\,Q_s^2(0)\,
 \frac{\rme^{(1+2 \sqrt{\abar}) Y}}{\sqrt{4\pi}\,(\sqrt{\abar}Y)^{3/2}}\,,\eeq
which shows that $\gamma_s\equiv 2 \sqrt{\abar}$ plays the role of an `anomalous'
saturation exponent within the fixed coupling scenario.

So far, we have used the tree--level definition of the saturation line,  $\bp^2=\hat q^{(0)}\tau$ or 
$\rho=Y$,  both in  the integration limit in \eqn{qren} and in the calculation
of the saturation momentum in \eqn{QsYFC}. As discussed in relation with \eqn{Qsx}, this choice
is ambiguous and the sensitivity of the results to this ambiguity may be viewed as an indication
of our error. To estimate this error, we can, for example, change the lower limit in each iteration 
with the updated value of $\hat{q}$, or even with the resummed value given in \eqn{qyy}. 
Both prescriptions lead to a correction of the same order, so let us follow the second one,
since it is rather easy to implement. 
Keeping only the leading asymptotic behavior of $\hat{q}_s(Y)$, as given by the exponential in \eqn{qyy}, we see that the lower limit of the $\rho_1$ integration in \eqn{qren2} changes from $Y_1$ to $(1+ 2\sqrt{\abar})Y_1$. Then it is an easy exercise to show that \eqn{qyrho} gets replaced by
 \beq
 \label{qyrholl2}
 \hat{q}(Y,\rho) =
 \hat{q}^{(0)}
 \bigg[
 \rmI_0\big(2 \sqrt{\abar Y \rho}\big) 
 - \frac{Y}{\rho}\,(1+ 2\sqrt{\abar})\,
 \rmI_2\big(2 \sqrt{\abar Y \rho} \big)
 \bigg],
 \eeq  
For consistency, when evaluating this expression on the saturation line, one should now use 
$\rho= \rho_s(Y)$, with $\rho_s(Y) \equiv (1+ 2\sqrt{\abar})Y$ (cf. \eqn{QsYFC}). Then the leading terms
in the asymptotic expansion cancel again between the two terms in \eqn{qyrholl2} and the net
result at large $Y$ is similar to that in \eqn{qyy}, except for the replacement of the exponential there by
  \beq
 \label{qyy2}
\rme^{2Y \sqrt{\abar(1+ 2\sqrt{\abar})} }\,\simeq\,
\rme^{2 \sqrt{\abar}\, Y (1+\sqrt{\abar})}\,=\,
\rme^{2 \sqrt{\abar}\, Y}\,\rme^{2\abar Y}\,,
% \left[1 + \mcal{O}(1/\sqrt{\abar}Y) \right],
 \eeq    
where we have also used $\sqrt{\abar}\ll 1$. As compared to \eqn{qyy},
the exponent in  \eqn{qyy2} includes an additional contribution $2\abar Y$, due to the change in the 
slope of the saturation line.
This new contribution represents a perturbative correction of $\order{\sqrt{\abar}}$ {\em to the exponent},
so the leading, exponential, behavior at large $Y$ is not modified. But if one is interested in
$\hat{q}_s(Y)$ itself, and not only in its logarithm, then this additional piece in the exponent matters to $\order{1}$
for any $Y \gtrsim 1/\abar$. In other terms, the {\em prefactor} of the leading exponential is unambiguously
given by our current approximations only so far as $\abar Y\ll 1$. This is in agreement with the fact that,
by working in the DLA, the single--logarithmic contributions have been systematically neglected. 
We conclude that the DLA is an accurate approximation for $\hat{q}_s(Y)$ only
within the window \eqref{rangeY}, but a good approximation for $\ln\hat{q}_s(Y)$ for arbitrary $Y\gtrsim 
1/\sqrt{\abar}$.

It is quite straightforward to generalize the previous discussion to the 
$\rho$--dependent initial condition\footnote{ Note that, as compared to \eqn{qhat02}, we ignore the 
constant shift $\rho_0$ in the value of $\rho$, since the corresponding effect can be trivially added:
the solution corresponding to an initial condition $\hat{q}^{(0)}(\rho) =
 \tilde{q}^{(0)}(\rho+\rho_0)$ is simply the sum of the solution \eqref{qyrhoic2} for $\hat{q}^{(0)}(\rho) =
 \tilde{q}^{(0)}\rho$ and that in \eqn{qyrho} with $\hat{q}^{(0)}=\tilde{q}^{(0)}\rho_0$.}
Then the pattern described below \eqn{q1} is still present and one similarly finds 
% that the $n$--th order contribution reads
% \beq
% \label{qnic2}
% \hat{q}^{(n)}(Y,\rho)=
% \tilde{q}^{(0)} \rho\, \frac{(\abar Y \rho)^n}{n! (n+1)!}
% -\tilde{q}^{(0)}(1-\delta_{n0})\, \frac{Y^2}{\rho}\,
% \frac{(\abar Y \rho)^n}{(n-1)!(n+2)!},
% \eeq  
%which is easily resummed to
 \beq
 \label{qyrhoic2}
 \hat{q}(Y,\rho) =
 \tilde{q}^{(0)}\frac{\rho}{\sqrt{\abar Y \rho}}
 \bigg[
 \rmI_1\big(2 \sqrt{\abar Y \rho}\big) 
 - \frac{Y^2}{\rho^2}\,\rmI_3\big(2 \sqrt{\abar Y \rho}\big)
 \bigg].
 \eeq
For $\rho=Y$, one can combine the two terms in the above to get
 \beq
 \label{qyyic2}
 \hat{q}(Y) = \tilde{q}^{(0)}\,
 \frac{2\, \rmI_2\big(2 \sqrt{\abar}\, Y\big)}{\abar Y}
 = \tilde{q}^{(0)}\, 
 \frac{2 \rme^{2 \sqrt{\abar}\, Y}}{\sqrt{4\pi \abar}\,(\sqrt{\abar}Y)^{3/2}}
 \left[1 + \mcal{O}(1/\sqrt{\abar}Y) \right],
 \eeq    
where the second equality is valid in the limit $Y \gg 1/\sqrt{\abar}$. Comparing Eqs.~\eqref{qyy} and \eqref{qyyic2}, 
we see that the $Y$--dependence of the asymptotic solution (including the leading prefactor) 
is not altered due to the change in the initial condition.

\section{\label{sec:rc}The asymptotic solution with running coupling}

In exact analogy to the fixed coupling case, we can construct a formal solution to the running coupling 
equation \eqref{qren2rc} via successive iterations. This allows us to express the solution $\hat{q}(Y,\rho)$
as an infinite series, similar to \eqn{qseries}, where however the individual terms with $n\ge 1$
are considerably more complicated than for fixed coupling. 
For not too large values of $n$, these terms can be efficiently
computed with the help of symbolic, computer--assisted, manipulations. This is possible
because the kernel in the integral equation \eqref{qren2rc} is simple enough to analytically perform
the integrations, at each step in the iteration procedure. However, we shall not be able
to deduce the analytic form of these terms for arbitrary values of $n$ and even less to explicitly resum 
the whole series. Still, through a semi--numerical procedure to be later described, we will be able
to deduce the asymptotic behavior of the series at large $Y$. 
As we shall also demonstrate, this dominant behavior is universal within
the class of initial conditions of interest. So, 
before we consider the more realistic initial condition in \eqn{qhat02}, let us first
assume the simple scenario in which $\hat{q}^{(0)}$ is momentum independent. Since here we are 
primarily interested in the asymptotic expansion, we can neglect $\rho_0$ in the denominator 
of the running coupling in \eqn{qren2rc}.
(This assumption will be relaxed in the numerical estimates in the following section.) We then
successively find
 \begin{align}
 &\hat{q}^{(1)}(Y,\rho) = 
 \hat{q}^{(0)}b  Y 
 \left(\ln \frac{\rho}{Y} +1 \right),
 \\
 &\hat{q}^{(2)}(Y,\rho) = 
 \hat{q}^{(0)}
 (b  Y)^2
 \left(\frac{1}{4}\ln^2 \frac{\rho}{Y} 
 + \frac{3}{4}\ln \frac{\rho}{Y} + 
 \frac{3}{8} \right),
 \\
 &\hat{q}^{(3)}(Y,\rho) = 
 \hat{q}^{(0)}
 (b  Y)^3
 \left(\frac{1}{36}\ln^3 \frac{\rho}{Y} +
 \frac{11}{72}\,\ln^2 \frac{\rho}{Y} 
 + \frac{49}{216} \ln \frac{\rho}{Y} + 
 \frac{49}{648} \right),
 \\
 & \cdots,
 \nn
 \label{qnrc}
 &\hat{q}^{(n)}(Y,\rho) = 
 \hat{q}^{(0)}
 (b  Y)^n 
 \left[\frac{1}{(n!)^2}\ln^n \frac{\rho}{Y} +
 c_{n,n-1}\,\ln^{n-1}\frac{\rho}{Y} 
 + \cdots +
 n c_{n,0} \ln \frac{\rho}{Y} + 
 c_{n,0} \right].
 \end{align}
 
A few observations are in order here. If there was not for the $Y$--dependent `saturation boundary' 
(the lower limit $Y_1$ in the integral over $\rho_1$ in \eqn{qren2rc}), we would have to introduce an infrared
cutoff at $\rho_0$ (or, equivalently, restore the `shift' $\rho_0$ in the denominator of the running coupling),
in order to avoid infrared singularities. Then one would easily find that, for a given $n$, 
only the term with the highest power $\ln^n(\rho/\rho_0)$ appears. As shown in the above equation, 
the respective coefficient reads $c_{n,n} = {1}/{(n!)^2}$. The corresponding series is straightforwardly 
resummed, since recognized as the Taylor expansion of the function $\rmI_0(x)$ with 
$x=2 \sqrt{\,b  Y \ln(\rho/\rho_0)}$. This is indeed the expected solution for the standard DLA
equation with running coupling \cite{DeRujula:1974rf,Kovchegov:2012mbw}.

However, the presence of the saturation boundary significantly modifies the problem, even more than in the
case of a fixed coupling. For a given $n$, we have to sum a polynomial of order $n$ in $\ln(\rho/Y)$ whose coefficients, in general, do not seem to be given by a simple analytic formula. As explicitly shown in \eqn{qnrc} we have been able to find only the coefficient of the leading term and the ratio, equal to $n$, of the linear and constant terms. We shall come later to a discussion of the information that we can infer from this last, simple, relation.

Consider now the value of this series along the (tree--level) saturation line,
$\rho=Y$. Then, for any $n$, we are left with only the constant term of the respective polynomial, namely
 \beq
 \label{qyyrc}
 \hat{q}_s(Y) = \hat{q}^{(0)}
 \sum_{n=0}^{\infty}
 c_n (b  Y)^n, 
 \eeq  
where we have written $c_n \equiv c_{n,0}$. The fact that all the logarithmic terms within
the polynomial $\hat{q}^{(n)}(Y,\rho)$ cancel for $\rho=Y$ implies that the asymptotic
behavior of the quantity $ \hat{q}_s(Y)$ at large $Y$ should be very different from that of the
standard DLA solution (with RC), and also much more difficult to obtain.
Indeed, as suggested by the first three iterations given explicitly above, 
it seems difficult to find a general analytic expression for the coefficients $c_n$, valid for arbitrary $n$. 
Nevertheless, one can explicitly construct these coefficients up to very high orders, via iterations,
by using a suitable mathematical program for symbolic manipulations. 
Then, as we shall see in a moment, 
the dominant behavior at large $n$ is $c_n\propto 1/(n!)^2$ (as for the evolution at fixed coupling). 
This in particular implies that, for a given $Y$, we only need a finite number of terms to reliably calculate $\hat{q}_s(Y)$ terms (cf.~\eqn{ncrc} below). More precisely, we shall argue 
that we need about $n \sim 2 \sqrt{b Y}$ terms. Vice versa, by keeping 
$n$ terms in the series, one can reliably calculate $\hat{q}_s(Y)$ up to rapidities 
$Y \sim n^2/4b $. In practice, one can easily check where to stop by requiring 
that adding an extra term in the series does not change the result for $\hat{q}_s(Y)$, to the desired
accuracy.

This discussion shows that, for any given $Y$, one can accurately compute the solution $\hat{q}_s(Y)$
by considering only a finite truncation of the series, whose coefficients are analytically known. This being said, 
it would be appealing to have a closed form of $\hat{q}_s(Y)$ in terms of known functions. As we now explain, 
this becomes feasible at sufficiently large $Y$. Namely, by numerically fitting the `$Y$--data', that is, the numerical 
values of $\hat{q}_s(Y)$ obtained from the properly truncated series for large values of $Y$, we have found the
following asymptotic expansion:
 \beq
 \label{dlogqrc}
 \frac{\dif \ln \hat{q}_s(Y)}{\bbar \dif Y} = 
 \frac{2}{\sqrt{\bbar Y}} - 
 \frac{2^{1/6}|\xi_1|}{(2\bbar Y)^{5/6}}+
 \frac{1}{4 \bbar Y} + \mcal{O}\big(Y^{-7/6}\big),
 \eeq
where $\xi_1 =-2.338 \dots$ is the rightmost zero of the Airy function. Via numerical tests, we have checked 
that the form of \eqn{dlogqrc}, including all the shown coefficients, is very robust: any tiny variation will not lead to a well--defined asymptotic series in which the remainder, here the final term of $\mcal{O}\big(Y^{-7/6}\big)$, remains smaller in magnitude than the leading terms\footnote{As an example of the kind
of tests that we performed, notice that $|\xi_1|$ is quite close to $7/3=2.333...$; however, if one replaces
$|\xi_1|\to 7/3$ in  \eqn{dlogqrc} and then one plots the function $Y^{7/6}\Delta(Y)$, with 
$\Delta(Y)$ the difference between the `exact' result for
$\ln \hat{q}_s(Y)$ (the numerical evaluation of the series truncated to high enough accuracy)
and the three explicit terms in its asymptotic expansion \eqref{dlogqrc}  with $|\xi_1|\to 7/3$, 
then one clearly sees that this function deviates from a constant and this deviation increases
with $Y$.}.

A heuristic, yet suggestive, way to understand the dominant term in \eqn{dlogqrc} 
(including its coefficient) is as follows: keeping only this leading term, one can rewrite \eqn{dlogqrc} as
 \beq
 \label{dlogqleading}
 \frac{\dif \ln \hat{q}_s(Y)}{\dif Y} = 
 2 \sqrt{\abar(Y)} + \cdots.
 \eeq
This is formally identical to the corresponding fixed coupling result, as extracted from \eqn{qyy}, 
and with the coupling in the latter evaluated at $\rho=Y$ (the natural scale indeed). A similar
relation between the asymptotic behaviors at fixed and respectively running coupling has also been
observed in the case of a shockwave \cite{Iancu:2002tr,Mueller:2002zm}. Perhaps even more remarkable,
and also quite intriguing, this resemblance with the corresponding shockwave problem extends to
the {\em second} term in the asymptotic expansion \eqref{dlogqrc}, i.e. the first preasymptotic term
$\propto Y^{-5/6}$. Indeed, exactly the same term appears in the asymptotic expansion of the 
logarithmic derivative of the saturation momentum $\dif \ln Q_s^2(Y)/\dif Y$ as obtained from the
BK equation (or from the BFKL equation with a saturation boundary) \cite{Mueller:2002zm,Munier:2003sj}. In fact,
it was this observation that has led us to `guess' the form of this particular term when trying to fit the $Y$--data. We shall comment in a while on the origin of the last term\footnote{Such a term seems to be absent in the asymptotic expansion of the $\dif \ln Q_s^2(Y)/\dif Y$, cf.~\cite{Beuf:2010aw}.} $\propto 1/Y$   in \eqn{dlogqrc}.       
 
The terms given in \eqn{dlogqrc} are sufficient\footnote{Two more preasymptotic terms could be possibly calculated along the lines of \cite{Beuf:2010aw}, but their contribution becomes irrelevant at large values of $Y$.}, 
modulo a constant arising from the integration, to determine the 
asymptotic behavior  of the jet quenching parameter at large $Y$~:
\beq
 \label{logqrc}
 \ln \hat{q}_s(Y)\,= \,4{\sqrt{\bbar Y}} - 
 {3|\xi_1|}{(4\bbar Y)^{1/6}}+
 \frac{1}{4}\ln Y +\kappa + \mcal{O}\big(Y^{-1/6}\big)\,.
 \eeq
We have numerically evaluated the additional constant term $\kappa$ 
in the asymptotic expansion of $\ln \hat{q}_s(Y)$
and found it to be close to 5.7. When exponentiated, this leads to a unnaturally large multiplicative coefficient, 
close to 300, in the expression for $\hat{q}_s(Y)$. This large factor finds its origin in the fact that we 
have neglected the constant $\rho_0$ when performing the transverse momentum integrations. Indeed, although these integrations are finite as they stand, there are strongly sensitivity to the lowest momenta (where the coupling
is stronger), thus generating very large contributions. 
As we shall verify in the next section, the inclusion of a realistic value for $\rho_0$ strongly
suppresses the magnitude of the solution $\hat{q}_s(Y)$, while leaving unchanged the asymptotic behavior
\eqref{dlogqrc} of the derivative of $\ln\hat{q}_s(Y)$.

 \eqn{logqrc} also shows that, with running coupling,  the medium saturation momentum
has the following dominant exponential behavior at large $Y$~:
\beq\label{QsYRC}
Q_s^2(Y)\,\propto\,
\rme^{Y+ 4\sqrt{\bbar Y}}\,.\eeq
As compared to the fixed coupling scenario, cf. \eqn{QsYFC}, the effects of the radiative corrections
are now milder: there is no `anomalous' contribution to the saturation exponent anymore, just a correction
to $\ln Q_s^2(Y)$ which grows like $\sqrt{Y}$. This is very similar to what happens in the case of a shockwave
\cite{Iancu:2002tr,Mueller:2002zm,Munier:2003sj}.

It is also useful to mention that one can equivalently fit the $n$--data, that is the coefficients $c_n$ in the expansion \eqn{qyyrc}. In such an approach one finds, for large $n$,
 \beq
 \label{cnfit}
 c_{n} = \tilde\kappa \,\frac{4^n \rme^{-3|\xi_1|n^{1/3}}}{(n-1)!n!},
 \eeq
where the proportionality factor is related to the constant  $\kappa$ introduced in \eqn{logqrc} 
via $\tilde\kappa=\sqrt{2\pi} \rme^{\kappa}$  and 
 can be numerically determined. Then, one can convert the summation in \eqn{qyyrc} into an integration which can be done by the steepest--descent method. For a given $Y$, the integrand is dominated by values of $n$ around
 \beq
 \label{ncrc}
 n_c = 2 \sqrt{\bbar Y} -\frac{|\xi_1|(4\bbar Y)^{1/6}}{2}.
 \eeq
This result justifies our previous statement concerning the numbers of terms needed in order to give an accurate result for the jet quenching parameter for a given $Y$. Moreover, one can see that the Gaussian integration around the saddle point $n_c$ leads to a prefactor proportional to $Y^{1/4}$ which in turn explains the third term in the expansion in equation \eqref{dlogqrc}. 

As in the fixed coupling scenario, the full asymptotic        
expansion given in \eqn{dlogqrc} does not seem to depend on the initial condition. We have found that \eqn{dlogqrc} remains valid when we consider either the initial condition of \eqn{qhat02}, or a similar one in which the logarithm in the numerator is absent.

Finally, we can get some information regarding the $\rho$--dependence of the two--variable function
$ \hat{q}(Y,\rho) $ for values of $\rho$ not too far from the saturation boundary. 
In this regime it is sufficient to add to the previous result only the term linear in $\ln(\rho/Y)$. 
This is rather easy to achieve, since the coefficients of the $\ln(\rho/Y)$ terms are related to 
those of the constant terms by just a factor of $n$ (cf.~\eqn{qnrc}). The analog of \eqn{qseries} becomes
 \beq
 \label{qyrhorc}
 \hat{q}(Y,\rho) = \hat{q}^{(0)}
 \sum_{n=0}^{\infty}
 c_n (b  Y)^n
 + \ln\frac{\rho}{Y}\,
 \hat{q}^{(0)}
 \sum_{n=1}^{\infty}
 n\, c_n (b  Y)^n + \cdots, 
 \eeq
where the dots stand for terms of higher order in $\ln(\rho/Y)$. Notice that to the order of accuracy one has that $\ln(\rho/Y) \simeq (\rho-Y)/Y$. It is trivial to see that the second term in \eqn{qyrhorc} can be expressed in terms of the $Y$--derivative of the first term [which is the quantity that we have called $\hat{q}_s(Y)$], so that 
 \beq
 \label{qyrhorc2}
 \hat{q}(Y,\rho)
 \simeq \hat{q}_s(Y) +(\rho-Y) \frac{\dif \hat{q}_s(Y)}{\dif Y} 
 \simeq \left(1+  \frac{2 (\rho-Y) \sqrt{\bbar}}{\sqrt{Y}} \right) 
 \hat{q}_s(Y).
 \eeq
The first equality in the above is general (it holds for both fixed or running coupling) : this is the
beginning of the Taylor expansion of  $\hat{q}(Y,\rho)$ near $\rho=Y$. Indeed, by inspection of 
the integration limits in Eqs.~\eqref{qren2} and \eqref{qren2rc}, one can verify that
\beq
\frac{\del \hat{q}(Y,\rho)}{\del Y}\Big |_{\rho=Y} =0\quad\Longrightarrow\quad
\frac{\del \hat{q}(Y,\rho)}{\del\rho}\Big |_{\rho=Y} = \frac{\dif \hat{q}_s(Y)}{\dif Y}\,.\eeq
 For the second equality in \eqn{qyrhorc2}, we have used the leading asymptotic term in \eqn{dlogqrc}
 (or, equivalently, \eqn{dlogqleading}).
The expansion in \eqn{qyrhorc} is valid so long as the 
second term, linear in the separation $\rho-Y$ from the saturation boundary, 
remains much smaller than the first one. For the running coupling case, this is the case provided  
$\rho-Y\lesssim \sqrt{Y/b }\sim{1/\sqrt{\abar(Y)}}$, which leaves us with a parametrically large 
window in which the $\rho$--dependence is indeed under control.

 \section{\label{sec:fy} Numerical studies and non--universal aspects}
 
In the previous sections, we have mostly focused on universal aspects of the evolution,
like the asymptotic behavior of the (renormalized) jet quenching parameter $\hat{q}_s(Y)$ for %sufficiently
large values of $Y$, which are insensitive to the details of the initial condition $ \hat{q}^{(0)}$,
such as the constant shift $\rho_0$ in the momentum variable $\rho$ in \eqn{qhat02}.
This has enabled us to analytically perform the energy ($Y$) and momentum ($\rho$) integrations
in the successive iterations of the evolution equation. 
In the case of a fixed coupling, this permitted us to deduce exact 
solutions for two different initial conditions, cf. Eqs.~\eqref{qyrho} and \eqref{qyrhoic2}. In the
running coupling case, the corresponding analysis allowed us to accurately determine 
(via a numerical fit to the truncated series obtained via iterations) the asymptotic behavior of  
$\ln \hat{q}_s(Y)$, up to terms which die away as $Y\to\infty$, cf. \eqn{logqrc}.

However, some interesting questions are left unanswered by the previous analysis,
among which, how fast is the approach towards asymptotics, what are the physical consequences
of the shift $\rho_0$, and what is the net effect of the quantum evolution (say, as measured by the enhancement
factor $\hat{q}_s(Y)/ \hat{q}^{(0)}$) for phenomenologically relevant values of $Y$ and $\rho_0$.
In this section, we shall try and answer such questions via semi--numerical studies, based
on appropriate truncations of the iterative series in which the individual terms are computed
analytically, but with the help of computer--assisted symbolic manipulations.
Such manipulations become more tedious when $\rho_0\ne 0$, which limits
the number of terms in the series that can be efficiently computed in that case. It is therefore important 
to have a good control of the convergency of the truncated series (for a given value of $Y$).
This will be first tested in the case of a fixed coupling, where it is possible to compare with
the exact respective results.

\begin{figure}
\begin{minipage}[b]{0.47\textwidth}
\begin{center}
\includegraphics[scale=0.5]{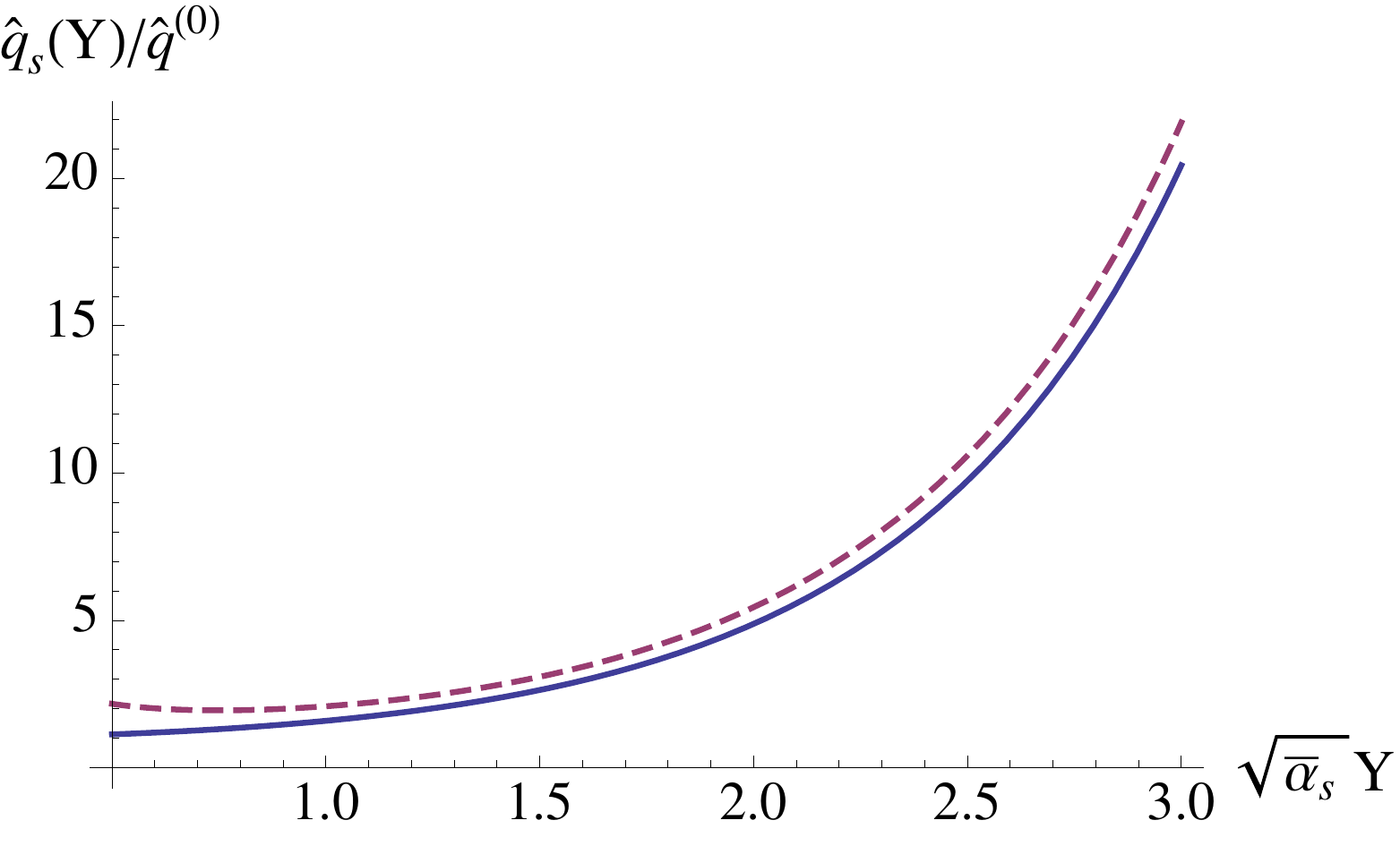}\\{\small (a)}
\end{center}
\end{minipage}
\hspace{0.04\textwidth}
\begin{minipage}[b]{0.47\textwidth}
\begin{center}
\includegraphics[scale=0.5]{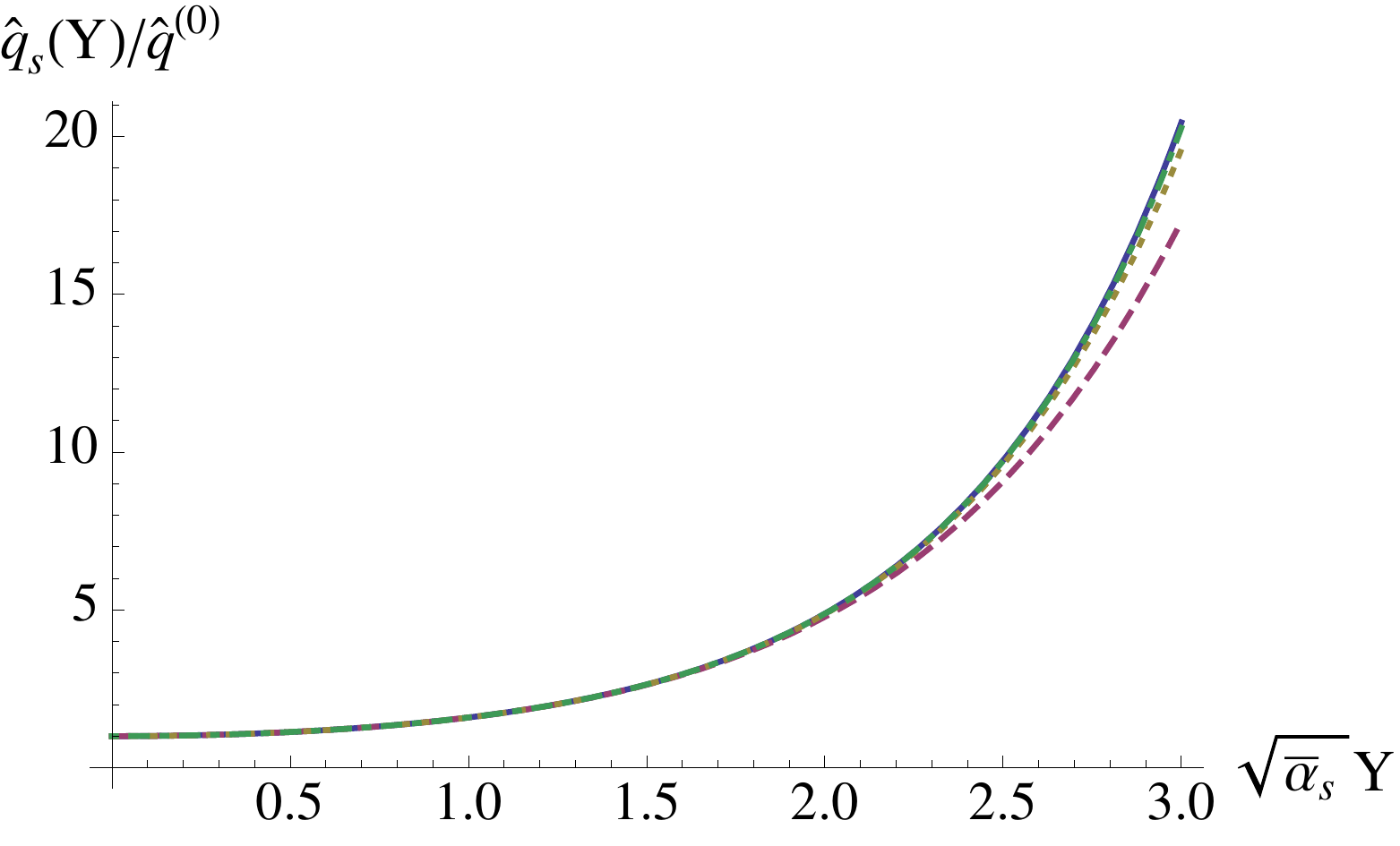}\\{\small (b)}
\end{center}
\end{minipage}
\caption{\sl The jet quenching parameter $\hat{q}_s(Y)$, normalized to its tree level value $\hat{q}^{(0)}$, as a function of $\sqrt{\abar} Y$ for fixed coupling evolution and a $\rho$--independent initial condition. (a) Continuous blue: exact solution. Dashed magenta: asymptotic solution.  (b) Continuous blue: exact solution. Dashed magenta, dotted gold, dotted--dashed green: truncated solutions with $n=3,4,5$ terms added to the tree level. The last curve, for $n=5$, cannot be distinguished from the exact solution.}
\label{fig:qhatfc}
\end{figure}

For definiteness, we consider the $\rho$--independent initial condition,
for which the exact solution with fixed coupling and its asymptotic expansion are both shown in \eqn{qyy}.
In Fig.~\ref{fig:qhatfc}.a we show a comparison between the exact and the (leading) asymptotic solution, 
for various values of the evolution `time' $\sqrt{\abar}Y$. 
As clear from this figure, the agreement is very good down to values $\sqrt{\abar} Y \simeq 1$. 
Then, we study the convergency of the truncated series. We would like to estimate the maximum value of $Y$ up to which such a truncated solution will be trustworthy. Via analytic considerations, similar to those
presented in the context of a running coupling (cf.~\eqn{ncrc}), one finds that the summation is dominated 
by values of $n$ around $n_c \sim \sqrt{\abar}Y$. Thus, it is not surprising that 
the $n=4$ truncated solution is in good agreement with the exact one up to values $\sqrt{\abar} Y \sim 3$,
as shown in Fig.~\ref{fig:qhatfc}.b. Similar features were observed long time ago in the solution to the fixed coupling BFKL equation \cite{Mueller:1986ey}. There, a comparison of the truncated solution (with $n=4 \div 5$) with the asymptotic one was done, and good agreement was found in a rather wide region of intermediate values of the evolution variable. To summarize, the analysis of the fixed coupling case in Fig.~\ref{fig:qhatfc} demonstrates
that, for a given $Y$, not only it is enough to keep a finite number of terms in the iterative solution, 
but also that the corresponding result agrees quite well with the asymptotic expansion already for 
relatively small values of $Y$ (and hence for a small number of terms in the truncated series).

\begin{figure}
\begin{minipage}[b]{0.47\textwidth}
\begin{center}
\includegraphics[scale=0.50]{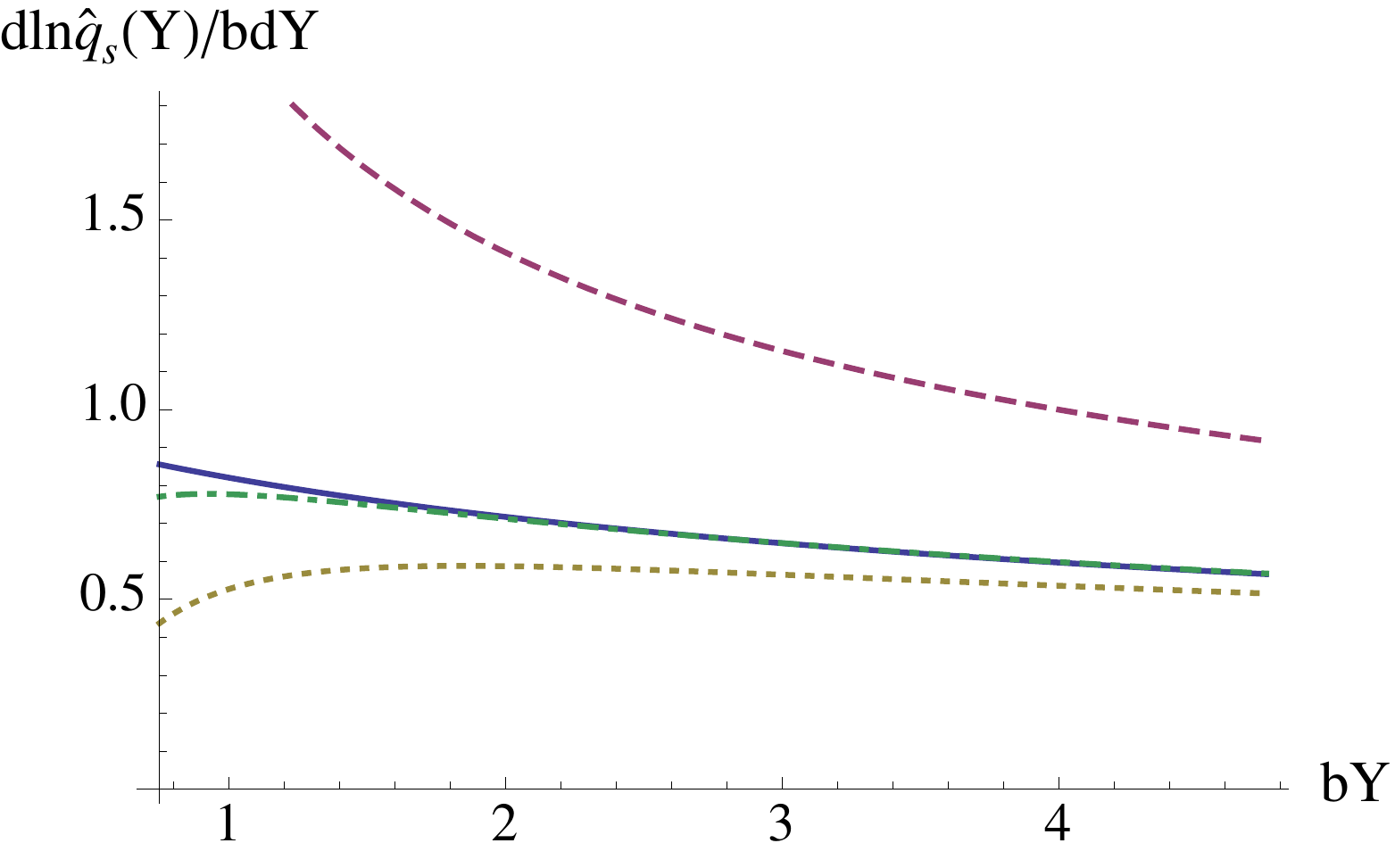}\\{\small (a)}
\end{center}
\end{minipage}
\hspace{0.04\textwidth}
\begin{minipage}[b]{0.47\textwidth}
\begin{center}
\includegraphics[scale=0.50]{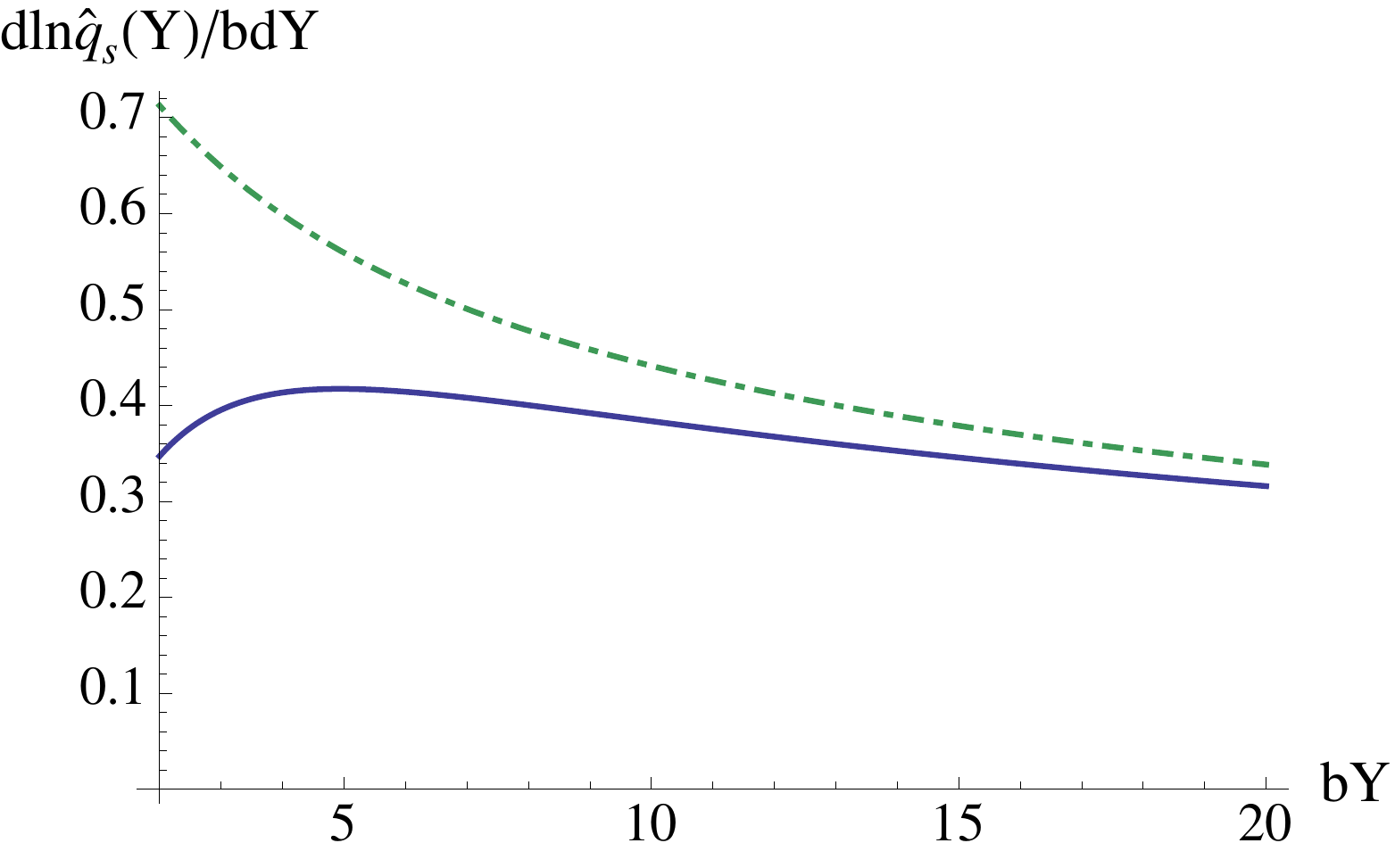}\\{\small (b)}
\end{center}
\end{minipage}
\caption{\sl The logarithmic derivative of the jet quenching parameter $\dif\ln\hat{q}(Y)/\dif Y$ for running coupling evolution as a function of $\bbar Y$. (a) $\rho$--independent initial condition.
Continuous blue: $n=7$ truncated solution
for $\rho_0=0$. Dashed magenta, dotted gold, dotted-dashed green: asymptotic solution given in \eqn{dlogqrc} with one, two, all terms kept respectively. (b) $\rho$--dependent initial condition, cf. \eqn{qhat02}  
with $\rho_0=2.3$ and $\bbar=4/3$. Continuous blue: $n=10$ truncated solution. Dotted-dashed green: asymptotic solution given in \eqn{dlogqrc} with all terms included.}
\label{fig:qhatrc}
\end{figure}

Turning to the case of a running coupling, we observe that, even though in that case we do not dispose
of an explicit solution in closed form, we can still check the convergence of the truncated solution, 
by comparing successive truncations with the each other. We have performed such
numerical tests and found that the estimate \eqn{ncrc} for the number of terms to be kept in the series
for a given value of $Y$, that is, $n \sim 2 \sqrt{b Y}$, is indeed reliable. Once again, the asymptotic
behavior is reached already for relatively small values of the corresponding evolution time $\bbar Y$.
This is illustrated in Fig.~\ref{fig:qhatrc}.a where we show the logarithmic derivative of the truncated solution for running coupling. We find that $n=6$ terms are enough to accurately reproduce the solution up to values of $\bbar Y$ close to 5, as adding more terms does not change the result. Furthermore, we see that the asymptotic solution in \eqn{dlogqrc} remains quite good down to $\bbar Y \sim 1$.

As discussed earlier, in the running coupling case one eventually needs to include a non--zero value for
the variable $\rho_0$ (cf.~Eqs.~\eqref{qren2rc} and \eqn{qhat02}), otherwise the evolution becomes extremely fast since sensitive to unphysically large values of the coupling.  At this level, it becomes appropriate to open
a parenthesis and discuss some physical choices for the quantities $\rho_0$ and $Y$. As a rough estimate
for the case of hot QCD matter (a weakly coupled QGP), let us use $\lambda=1/T$ with $T=500$~MeV and
$L=3\div 8$~fm; this yields $Y=\ln (LT) \simeq 2\div 3$. Also, to obtain an estimate for
$\rho_0 = \ln\big(\hat{q}^{(0)} \lambda/\Lambda^2\big)$, we also need the tree--level
jet quenching parameter. Taking (once again, as a very rough estimate) $\hat{q}^{(0)} =1\ {\rm GeV}^2/$fm
together with $\Lambda=200$~MeV, one finds $\rho_0\simeq \ln 10 \simeq 2.3$.
Closing the parenthesis and returning to Fig.~\ref{fig:qhatrc}.b, we notice that, by including a non--zero value 
$\rho_0=2.3$ in the calculation, one does not alter the asymptotic behavior of $\ln \hat{q}_s(Y)$,
as shown in \eqn{dlogqrc} (albeit the approach towards asymptotics appears to be slightly slower than
for $\rho_0=0$, cf. Fig.~\ref{fig:qhatrc}.a).

\begin{figure}
\begin{minipage}[b]{0.47\textwidth}
\begin{center}
\includegraphics[scale=0.50]{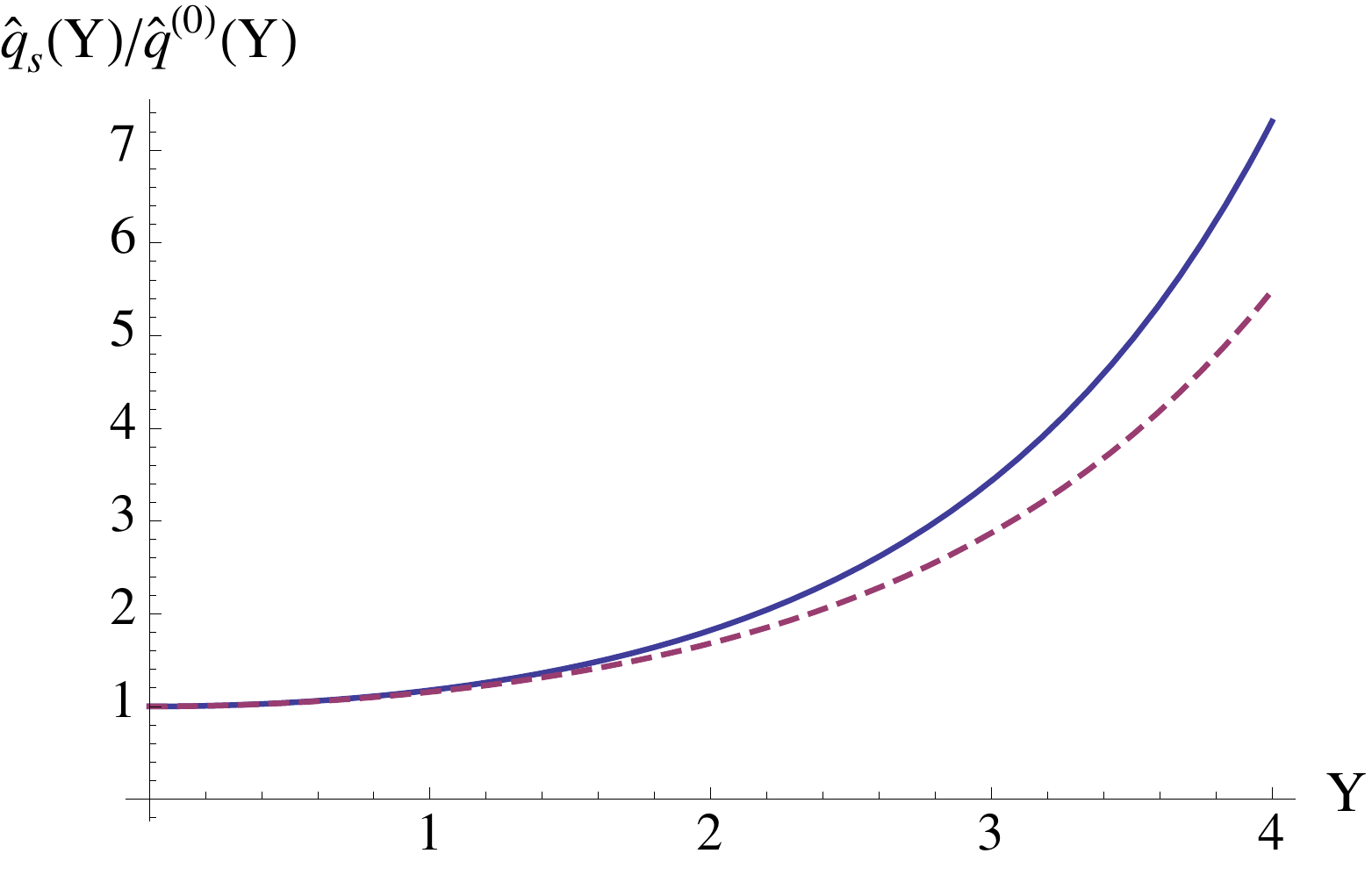}\\{\small (a)}
\end{center}
\end{minipage}
\hspace{0.04\textwidth}
\begin{minipage}[b]{0.47\textwidth}
\begin{center}
\includegraphics[scale=0.50]{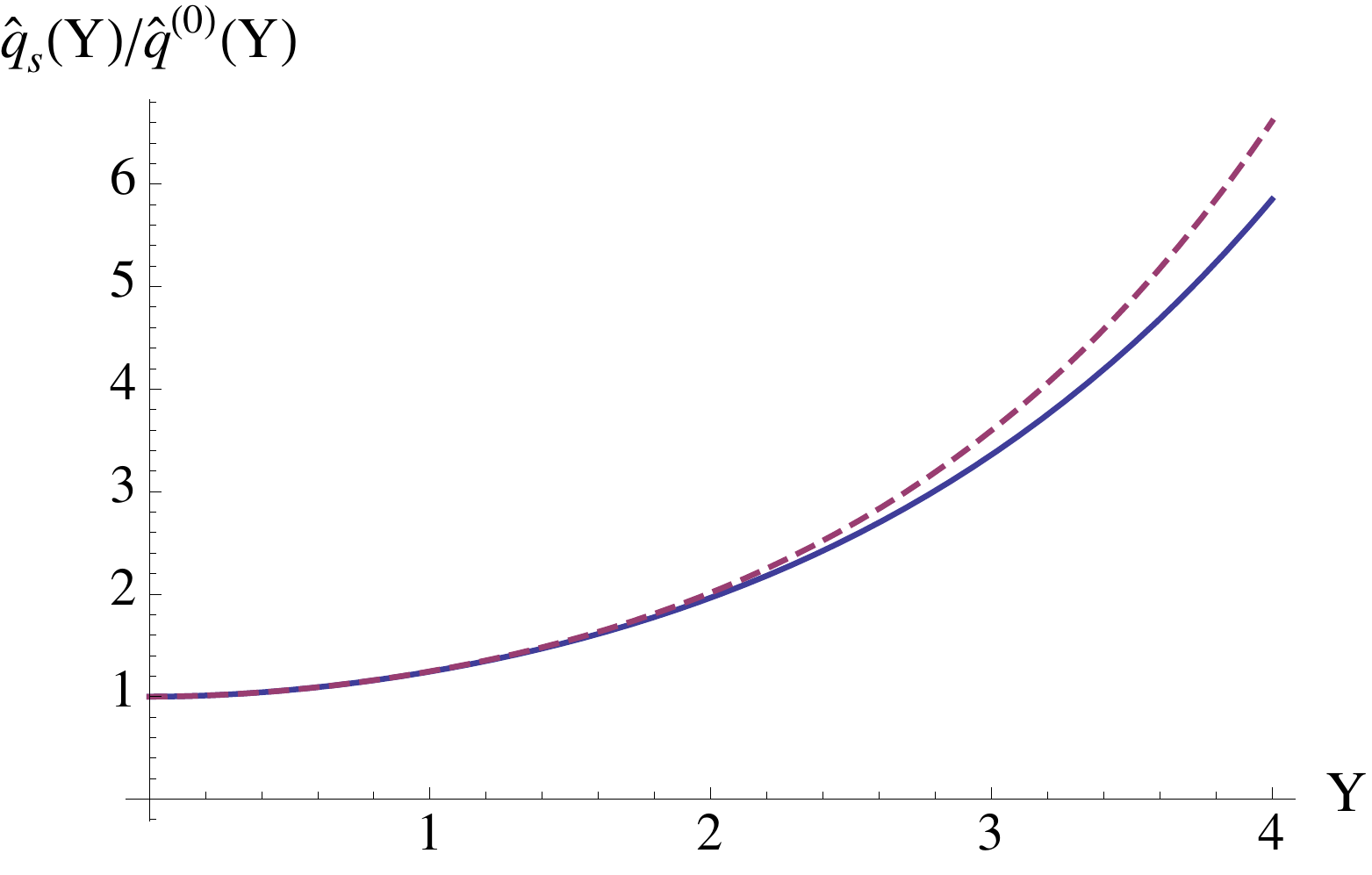}\\{\small (b)}
\end{center}
\end{minipage}
\caption{\sl Enhancement factor for the jet quenching parameter $\hat{q}$ as a function of $Y$. (a) Fixed coupling evolution with $\abar =0.33$. Continuous blue: $\rho$--independent initial condition. Dashed magenta: initial condition proportional to $\rho+\rho_0$ with $\rho_0=2.3$ and $\rho=Y$. 
(b) Running coupling evolution with $\bbar=4/3$ and $\rho_0=2.3$. Continuous blue: 
$\rho$--independent initial condition. Dashed magenta: initial condition proportional to $\ln(\rho+\rho_0)/(\rho+\rho_0)$ with $\rho_0=2.3$ and $\rho=Y$. At $Y=3$ and for constant initial condition the enhancement factor is 3.4 for both types of evolution. For the $\rho$--dependent initial condition the corresponding factor is 2.9 for fixed coupling and 3.6 for running coupling.}
\label{fig:qhat}
\end{figure}

Finally, in Fig.~\ref{fig:qhat} we show the enhancement factor $\hat{q}_s(Y)/ \hat{q}^{(0)}(Y)$ for both
fixed and running coupling and for two different types of initial conditions. 
(The function  $\hat{q}^{(0)}(Y)$ is the respective
initial condition in \eqn{qhat02} evaluated at $\rho=Y$.)
As visible in these figures, there is roughly an enhancement factor $3\div 4$ in the value of the jet quenching parameter after after a quantum evolution of three units in rapidity, for both fixed and running coupling.
(To estimate the uncertainty coming from the choice of the shift $\rho_0$, we vary the latter in between 1.8 and 
2.8. We thus find that, in the RC case, the enhancement factor varies from 3.9 to 3.0,
whereas with  FC, it is almost constant and approximately equal to 2.9.) 
The fact that this factor appears to be similar with both fixed 
and running coupling is likely to be `accidental', in the sense that the respective predictions will start 
deviating from each other for larger values of $Y$. To render this manifest, we compare on a
same plot, in Fig.~\ref{fig:qhat5}, the enhancements factors corresponding to both 
fixed and running coupling (with constant initial conditions, for simplicity), for values of
$Y$ which are only slightly larger than those in Fig.~\ref{fig:qhat}. Whereas the two curves
closely overlap up to $Y=3$ (in agreement with Fig.~\ref{fig:qhat}), 
they differ by a factor of 2 when $Y=5$. With further increasing $Y$, this deviation is rapidly 
growing.

%\vspace*{0.5cm}
%\noindent \hrulefill \qquad 
%END OF CORRECTIONS 
%\qquad \hrulefill
%\vspace*{-.6cm}

\begin{figure}
\begin{center}
\includegraphics[scale=0.60]{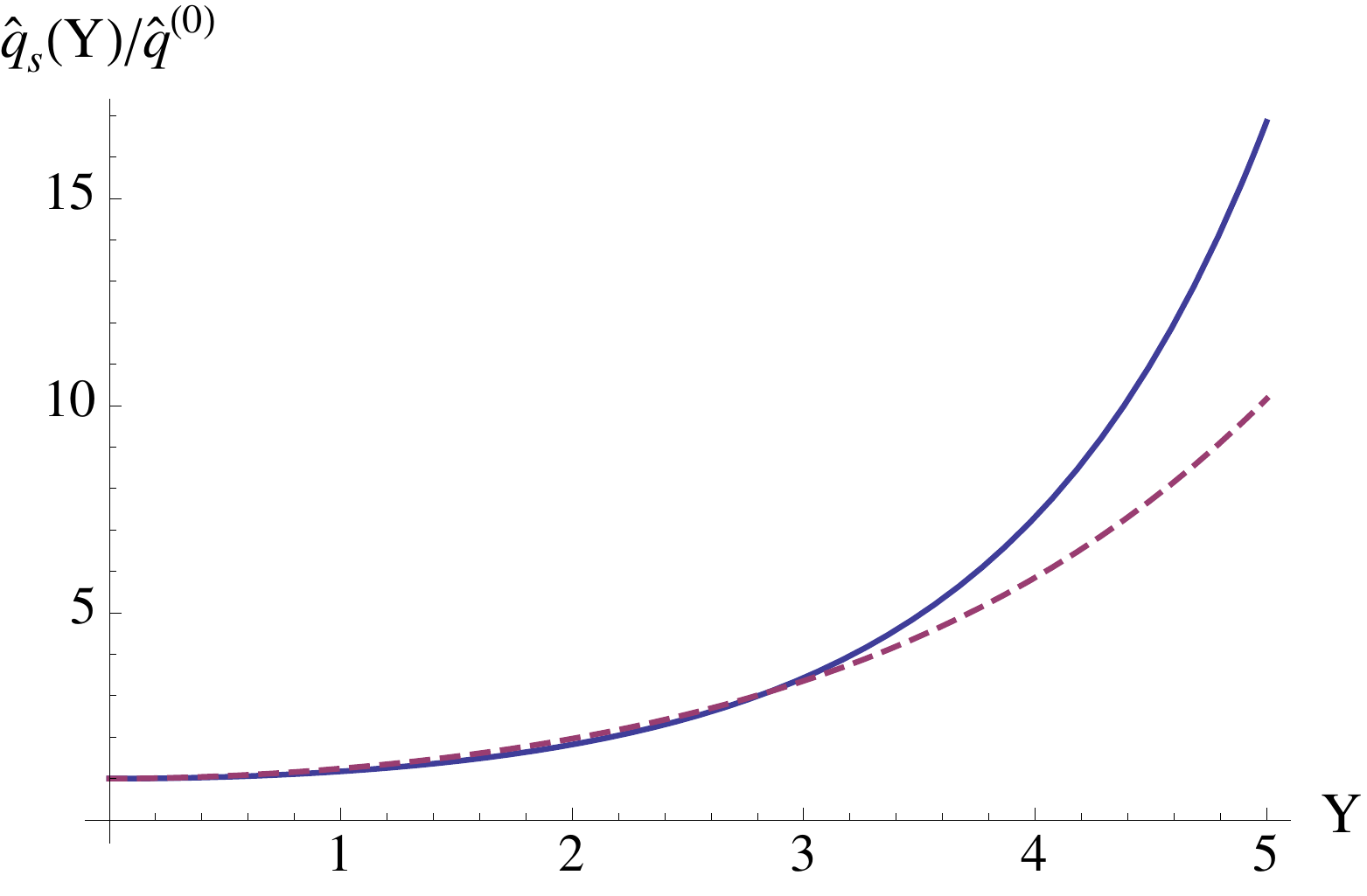}
\end{center}
\caption{\sl Enhancement factor for the jet quenching parameter $\hat{q}$ as a function of $Y$, for
both fixed and running coupling and with $\rho$--independent initial conditions. 
Continuous blue: 
fixed coupling evolution with $\abar =0.33$. Dashed magenta: running coupling evolution with 
$\bbar=4/3$ and $\rho_0=2.3$.}
\label{fig:qhat5}
\end{figure}

%\newpage
\begin{acknowledgments}
%Acknowledgments}
The research is supported by the Agence Nationale de la Recherche under the project \# 11-BS04-015-01 and by the European Research Council under the Advanced Investigator Grant ERC-AD-267258. Fig.~\ref{fig:evol} has been created with Jaxodraw \cite{Binosi:2003yf}.
\end{acknowledgments}

%\appendix*
%\section{\label{sec:app}To be written}

%\bibliographystyle{utcaps}
%\bibliography{../../References/refs}

\providecommand{\href}[2]{#2}\begingroup\raggedright\endgroup

\end{document}